\documentclass[aps,prl,twocolumn,amsmath,amssymb,nofootinbib,superscriptaddress,floatfix]{revtex4-1}

\usepackage{ulem}
\usepackage{amsmath,amsfonts,amssymb}
\usepackage{amssymb}
\usepackage{amsthm}
\usepackage[dvipdf]{color}
\usepackage{graphicx}
\usepackage{dcolumn}
\usepackage{bm} 
\usepackage[rightcaption]{sidecap}

\begin{document}

\title{Nonlinear Topological Mechanics in Elliptically Geared Isostatic Metamaterials}

\author{Fangyuan Ma}
\thanks{These authors contribute equally to this work.}
\affiliation{Key Lab of Advanced Optoelectronic Quantum Architecture and Measurement (MOE), School of Physics, Beijing Institute of Technology, Beijing, 100081, China}

\author{Zheng Tang}
\thanks{These authors contribute equally to this work.}
\affiliation{Key Lab of Advanced Optoelectronic Quantum Architecture and Measurement (MOE), School of Physics, Beijing Institute of Technology, Beijing, 100081, China}

\author{Xiaotian Shi}
\affiliation{Aeronautics and Astronautics, University of Washington, Seattle, Washington, 98195, USA}

\author{Ying Wu}
\affiliation{School of Science, Nanjing University of Science and Technology, Nanjing, 210094, China}

\author{Jinkyu Yang}
\affiliation{Aeronautics and Astronautics, University of Washington, Seattle, Washington, 98195, USA}
\affiliation{Mechanical Engineering, Seoul National University, Seoul, 08826, Korea}

\author{Di Zhou}
\email{dizhou@bit.edu.cn}
\affiliation{Key Lab of Advanced Optoelectronic Quantum Architecture and Measurement (MOE), School of Physics, Beijing Institute of Technology, Beijing, 100081, China}

\author{Yugui Yao}
\affiliation{Key Lab of Advanced Optoelectronic Quantum Architecture and Measurement (MOE), School of Physics, Beijing Institute of Technology, Beijing, 100081, China}

\author{Feng Li}
\email{phlifeng@bit.edu.cn}
\affiliation{Key Lab of Advanced Optoelectronic Quantum Architecture and Measurement (MOE), School of Physics, Beijing Institute of Technology, Beijing, 100081, China}

\begin{abstract}
Despite the extensive studies of topological systems, the experimental characterizations of strongly nonlinear topological phases have been lagging. To address this shortcoming, we design and build elliptically geared isostatic metamaterials. Their nonlinear topological transitions can be realized by collective soliton motions, which stem from the transition of nonlinear Berry phase. Endowed by the intrinsic nonlinear topological mechanics, surface polar elasticity and dislocation-bound zero modes can be created or annihilated as the topological polarization reverses orientation. Our approach integrates topological physics with strongly nonlinear mechanics and promises multi-phase structures at the micro and macro scales.
\end{abstract}

\maketitle
\emph{Introduction---}Since the discovery of topological insulators~\cite{PhysRevLett.61.2015,PhysRevLett.98.106803,PhysRevLett.42.1698}, an increasing interest in topological states has been addressed by the condensed matter community. Topological band theory has proliferated across classical structures, such as topological photonics~\cite{smirnova2020nonlinear, RevModPhys.91.015006, xia2021nonlinear, gangaraj2020topological, bomantara2017nonlinear, tuloup2020nonlinearity, abrashuly2021photonic}, electrical circuits~\cite{kotwal2021active, tang2023strongly, PhysRevLett.123.053902, hohmann2023observation}, acoustics~\cite{li2018weyl,luo2021observation,susstrunk2015observation,PhysRevLett.122.128001}, plasmonics~\cite{fu2021topological,fu2022topological} and mechanics~\cite{nash2015topological,kane2014topological,zhou2018topological,PhysRevX.9.021054,paulose2015topological,liu2021topological,PhysRevLett.116.135501,scheibner2020non,PhysRevResearch.2.023173}, which exhibit unconventional boundary responses. Most works are hitherto limited to linear regime, whereas the study of nonlinear topological systems remains sporadic.

To date, nonlinear topological metamaterials have been investigated in Kerr-nonlinear photonics~\cite{smirnova2020nonlinear,RevModPhys.91.015006,xia2021nonlinear,bomantara2017nonlinear,tuloup2020nonlinearity}, weakly nonlinear electrical~\cite{wang2019topologically} and mechanical systems~\cite{PhysRevE.97.032209,zhou2021amplitude,PhysRevB.101.104106}. Despite the efficacy of the superficial Kerr-nonlinear topological invariant inherited from linear theories, it remains unclear whether these weakly nonlinear excitations remain topological for larger amplitudes~\cite{bomantara2017nonlinear,tuloup2020nonlinearity,zhou2021amplitude,PhysRevB.101.104106,PhysRevE.95.022202,chen2014nonlinear,PhysRevLett.127.076802}. The symmetry-violating nonlinearities may deteriorate the topological robustness by causing mode instabilities~\cite{vakakis2001normal,PhysRevB.103.024106} and frequencies mixing with bulk bands~\cite{bomantara2017nonlinear,tuloup2020nonlinearity,PhysRevB.104.174306}. On the other hand, nonlinear excitations also provide unique features absent in linear systems, such as non-reciprocal phase transition~\cite{PhysRevE.97.032209}, moving domain walls~\cite{PhysRevLett.127.076802,PhysRevResearch.4.023211}, and transporting mechanical states~\cite{zhou2021amplitude,PhysRevE.95.022202}. However, the rigorous experimental demonstrations of strongly nonlinear topological phases and properties, are yet elusive~\cite{zhou2022NC}.

Herein, we invoke and experimentally demonstrate strongly nonlinear topological transitions in a mechanical prototype by assembling elliptic gears on beams. Recent advances in mechanical metamaterials~\cite{vitelliPRX,Fang2022NM} have showcased functionalities of \emph{circular} gears, including shape morphing and topological mechanics in the linear elastic regime. In contrast, our work exploits the geometric nonlinearity of \emph{elliptic} gears to uncover unprecedented physical phenomena when topology encounters built-in nonlinearity. In the elliptically geared one-dimensional (1D) system, the topological index~\cite{zhou2022NC} called quantized nonlinear Berry phase guarantees the nonlinear topological modes, and the geared lattice exhibits soliton-induced topological transition.

We use 1D geared metabeams to construct highly adjustable topological metamaterials in 2D. These metamaterials can undergo phase transitions in their topologically polarized mechanical properties, which are enabled by the interplay between the nonlinear geometric transition in the metabeams, and the soft shearing of the whole structure, called the Guest mode~\cite{guest2003determinacy}. While Guest modes are always nonlinear, their physical consequences are remarkably distinct from the nonlinear geometry of metabeams, which endows the topological phase transition of the metabeam mechanics, and are described by the nonlinear topological index. The term ``highly flexible topological metamaterials in 2D" specifically refers to the intrinsic nonlinear topological mechanics embedded in every metabeam, whereas Guest modes simply link them to the global topological polarization of the whole structure.
The geometric interplay between Guest modes and nonlinear topological transition in metabeams, allows for a complete switch in position between surface softness and rigidity as the topological polarization reverses its direction, which is a significant departure from previous research that only observed partial exchanges of topological mode localization~\cite{vitelliPRX}. Within our metamaterial, topological floppy modes or states of self-stress can be positioned near dislocations, and the bounded softness or rigidity can be annihilated or created as the topological polarization reverses direction. The reversal of the polarization is induced by the Guest mode, which integrates nonlinear topological mechanical transitions in every metabeam.

\begin{figure}[htbp]
\centering
\includegraphics[scale=0.78]{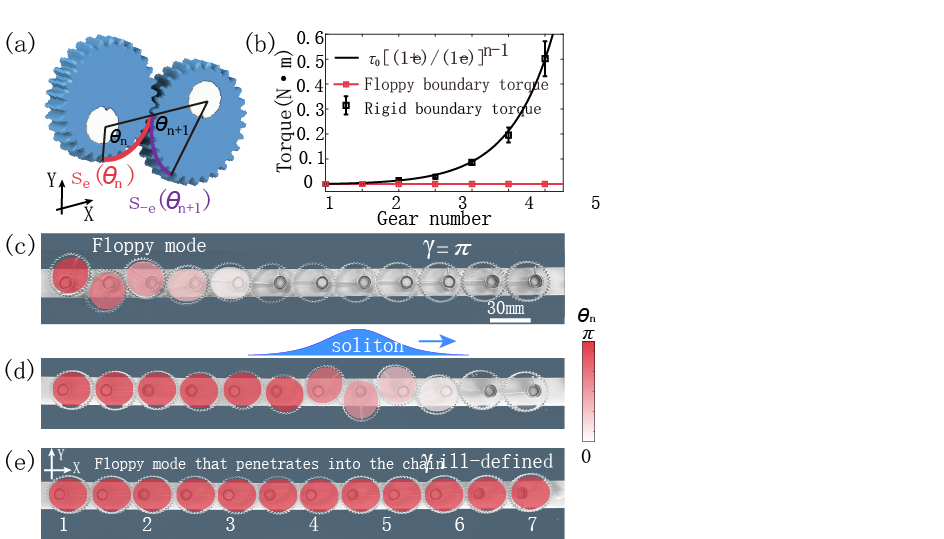}
\caption{Elliptically geared chain and its mechanical properties. (a) The gear parameters: thickness 5 mm, major axis $2a=30$ mm, minor axis $2b=27.5$ 
 mm, eccentricity $e=0.4$, number of teeth 21, and number of gears $N=12$. (b) Torque measurements of the gears at floppy and rigid boundaries. (c) The floppy mode localized at the left boundary. (d) A soliton that penetrates through the chain. (e) The end state of soliton propagation is that all gears rotate uniformly by $\pi$. This indicates a shift in the localization edge of the nonlinear topological floppy mode, where the left boundary becomes rigid and the right boundary becomes soft. 
}\label{fig1}
\end{figure}

\emph{Strongly nonlinear topological floppy modes in 1D geared chains---}Our prototype consists of a chain of elliptic gears coupled to their nearest neighbors, which are 3D-printed using photosensitive resin (Fig.\ref{fig1}(a)). Every gear can rotate freely about pivots on their right-sided focal points, and the nearest-neighbor gears keep engaged during rotations. The rotation angles of the gears are re-written in an alternative way $\theta_n\to(-1)^n \theta_n$ for the remainder of this letter. The elastic energy of the chain is expressed as $V=\sum_n k\ell_n^2 /2$, where $k$ is the elastic constant and $\ell_n$ is the sliding distance between adjacent gears, which reflects the teeth deformation. The sliding distance is given by
$\ell_n=s_e (\theta_n )-s_{-e} (\theta_{n+1} ),$
where $s_e(\theta)$ is the arc length of the contacting point traveling along the ellipse with eccentricity $e$ and rotation angle $\theta$ (see Supplementary Information~\cite{gearSI}). We define the degree of nonlinearity as $|s_e(\theta)/\theta a(1-e)-1|$, where $a$ denotes the major semi-axis of the ellipse. Circular gears with $e=0$ or small rotation angles with $\theta \approx 0$ lead to purely linear arc length in $\theta$, resulting in a vanishing degree of nonlinearity that demonstrates the linear elastic regime~\cite{vitelliPRX}. However, for gears with large eccentricity, like $e=0.4$ and rotation angles close to $\theta\approx\pi$, as shown in Fig.\ref{fig1}(c), the degree of nonlinearity reaches $0.6$, which manifests \emph{strong nonlinearity} in the gear mechanics.

Gear rotations typically induce both elastic deformation and rotational kinetic energy. However, floppy modes refer to zero-frequency angular displacements that do not deform elastic bonds or gear teeth. Therefore, all $\ell_n$ must be zero, resulting in zero potential energy. Finally, since floppy modes occur very slowly, their zero-frequency static nature means that the contribution of kinetic energy is also negligible.


The topological nature~\cite{kane2014topological,vitelliPRX} of the floppy modes is understood through the chain mechanics under periodic boundary condition (PBC). In the linear elastic regime, the mechanical properties are described by a gapped two-band model~\cite{PhysRevLett.42.1698,kane2014topological}. As the amplitudes increase, the frequencies of plane-wave nonlinear traveling modes deviate from Bloch waves, leading to a decrease in the nonlinear bandgap. At the critical amplitude $A_c=\pi$, a floppy mode penetrates into the chain, as shown in Fig.\ref{fig1}(e), which has a zero frequency indicating the closure of the nonlinear bandgap~\cite{zhou2022NC}. Thus, we define the amplitude-controlled topological invariant, namely quantized nonlinear Berry phase~\cite{zhou2022NC}, as
\begin{eqnarray}\label{2}
\gamma(A)=\pi[1+{\rm sgn}(e) \, {\rm sgn}(A_c-A)]/2,
\end{eqnarray}
where $A$ is the amplitude of plane-wave nonlinear traveling waves. This index distinguishes between different topological phases below and above $A_c=\pi$, reflecting topologically distinct nonlinear responses on the open boundaries of the lattice.

In the geared metamaterial shown by Figs.\ref{fig1}(c-e), open boundary conditions (OBCs) are used, severing the connection on the periodic boundary to create an excess degree of freedom, which is manifested as the floppy mode. In Fig.\ref{fig1}(c), the left boundary exhibits a nonlinear topological floppy mode when $\theta_1 < A_c$, reflecting a topologically-polarized metabeam with $\gamma(A<A_c)=\pi$. This topological polarization can be reversed by activating a soliton that propagates through the chain (Fig.\ref{fig1}(d)). When all gear rotations approach $A_c= \pi$ in Fig.\ref{fig1}(e), $\gamma(A=A_c )$ becomes ill-defined and induces topological phase transition. The floppy mode with gear rotations beyond $\pi$ is localized at the right boundary. Floppy modes account for the local stiffness of lattice boundaries, as evidenced by the highly asymmetric rigidity in Fig.\ref{fig1}(b).

The topological transition amplitude can be customized using other gear shapes, such as the triangle-trefoil geared chain~\cite{gearSI}, whose transition amplitude $A_c=\pi/3$ stems from its $C_3$-rotational symmetry. The idea of metabeams combines topological mechanics and strongly nonlinear transitions into a single nonlinear topological index, which goes beyond linear and weakly nonlinear topological mechanics~\cite{kane2014topological, vitelliPRX, PhysRevE.95.022202, PhysRevB.101.104106}. Additionally, the compact and highly tunable designs allow for the assembly of higher-dimensional mechanical metamaterials and customizes their rich topological mechanical phases.

\begin{SCfigure}[0.4][htbp]
\centering
\includegraphics[scale=0.6]{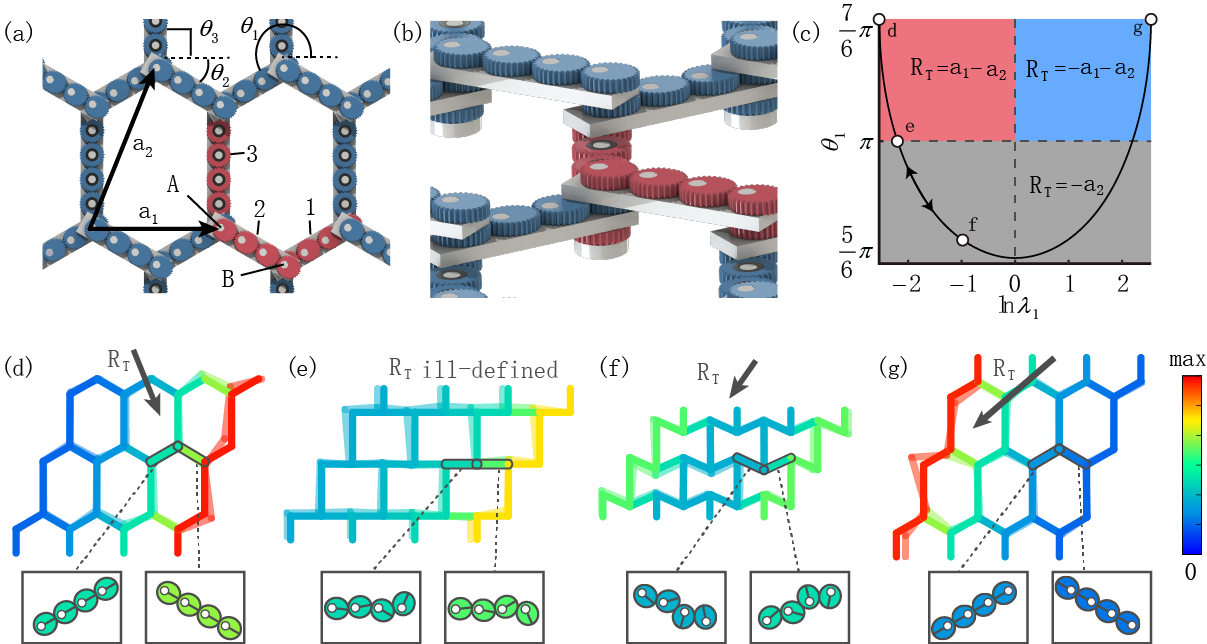}
\caption{Topological phases of the 2D geared metamaterial. (a), (b) Top and side views of the unit cell. (c) The multi-phase diagram of the lattice topology, under the parameters constraints of $\theta_2=\pi-\theta_1$ and $\lambda_2=\lambda_1^{-1}$ in (a). Three identified phases are characterized by $\bm{R}_{\rm T}$ and represented by different colored regions. White dots correspond to lattice configurations shown in panels (d-g). (d-g), Geometric configurations of the 2D lattices, where the topological polarizations are $\bm{R}_{\rm T}=\bm{a}_1-\bm{a}_2$ in (d), ill-defined $\bm{R}_{\rm T}$ in (e), $\bm{R}_{\rm T}=-\bm{a}_2$ in (f), and $\bm{R}_{\rm T}=-\bm{a}_1-\bm{a}_2$ in (g), respectively.}
\label{fig2}
\end{SCfigure}

\emph{Geared topological metamaterials in 2D---}We utilize the 1D prototype of elliptically geared metabeams to construct 2D highly flexible metamaterials in a generalized honeycomb lattice. In Fig.\ref{fig2}(a), three types of metabeams with the initial orientations $\theta_1=7\pi/6$, $\theta_2=-\pi/6$, and $\theta_3=\pi/2$, are prepared by assembling $N_1=4$, $N_2=4$, and $N_3=6$ elliptic gears on top of them, where the gear eccentricities are $e_1=e_2=0.4$, and $e_3=0$ (circular gears), respectively. The transmission rates, denoting the rotational speed ratio of gears at the last site to the first, are $\lambda_1=1/\lambda_2=12.7$ and $\lambda_3=1$ for the initial configurations of the metabeams. The primitive vectors $\bm{a}_1$, $\bm{a}_2$ and reciprocal vectors $\bm{b}_1$, $\bm{b}_2$ satisfy $\bm{a}_i\cdot \bm{b}_j=2\pi \delta_{ij}$. The above parameters are chosen for experimental convenience (see \cite{gearSI} for general choices).

The unit cell in Fig.\ref{fig2}(b) comprises three metabeams joined at a vertical hinge that penetrates through three co-axial gears, preventing relative displacements and rotations (see Fig.S10 for experimental manufacturing details~\cite{gearSI}). Each site features two translational and one rotational degrees of freedom ($N_{\rm DOF}=3$), and every metabeam offers one longitudinal constraint and one transverse constraint provided by the sliding distance ($N_{\rm con}=2$). The coordination number of the geared honeycomb lattice is $z=2N_{\rm DOF}/N_{\rm con}=3$, which results in a mechanical frame with the balanced degrees of freedom and constraints, ensuring it to stay at the isostatic point.

\widetext

\endwidetext

The mechanical properties are described by the compatibility matrix $\bm{C}$, which maps the displacements and rotations of the gears at each site, to the elongations and sliding distances of the beams. In the analysis of floppy modes, both kinetic and potential energy are negligible. Thus, floppy modes constitute the null space of $\bm{C}$, indicating that all beams and gear teeth remain undeformed. Self-stress states, which are characterized by the null space of $\bm{C}^\top$ ($\top$ denotes matrix transpose), describe non-zero tensions that allow for vanishing net forces and torques on each gear. Spatially repetitive frames have the property that for every wavevector $\bm{k}$ in Brillouin zone, the mechanical properties are governed by the Fourier-transformed compatibility matrix $\bm{C}(\bm{k})$. The topological mechanical phase is described by the winding numbers  
\begin{eqnarray}\label{3}
\mathcal{N}_i=-\frac{1}{2\pi}\oint_{C_i}d\bm{k}\cdot \nabla_{\bm{ k}} \, {\rm Im}\, \ln \det \bm{C}(\bm{ k}), \quad i=1,2,
\end{eqnarray}
where $C_i=\bm{ k}\to \bm{ k}+\bm{ b}_i$ is a closed-loop trajectory in reciprocal space. The winding numbers in the generalized honeycomb lattice remain invariant under arbitrary choices of $C_i$ due to the fully-gapped phonon spectra, except for the $\bm{k}=0$ point~\cite{kane2014topological}. These well-defined invariants constitute the vector 
$\bm{R}_{\rm T}=\sum_{i=1,2}\mathcal{N}_i\bm{a}_i, $
called the topological polarization, which characterizes the topological phases of the mechanical metamaterial.

Isostatic lattices can host nonlinear and uniform soft strains of the whole structure, known as Guest modes, that reversibly evolve the geometry and change the topological polarization without causing any elastic energy. To mark the rotation angle of the Guest mode, we use the bond orientation of the first-type metabeams, denoted as $\theta_1$. However, the gears also play a role by inducing a degree of nonlinearity within the metabeams, which can be quantified by the transmission rate $\lambda_1$~\cite{gearSI}. The topological polarization changes correspondingly, and follows the trajectory shown in the multi-phase diagram of Fig.\ref{fig2}(c). In~\cite{gearSI}, we also plot the topological phase diagram in terms of the Guest mode angle and degree of nonlinearity. Figs.\ref{fig2}(d-g) depict four lattice configurations that evolve continuously from Fig.\ref{fig2}(a) via the Guest mode, with corresponding gear orientations and topological polarizations displayed. We note that the elasticity analysis in the 2D metamaterial is based on a compatibility formalism that assumes small displacements from a reference configuration determined by the degree of the nonlinear Guest mode. The band structure of the compatibility formalism is based on linear elastic theory.

Bulk-boundary correspondence states that topological polarization reveals the localization of floppy modes on open boundaries. The number density of floppy modes per supercell is $\nu=\frac{1}{2\pi} \bm{G}\cdot (\bm{R}_{\rm T}+\bm{R}_{\rm L})$, where $\bm{G}$ denotes the reciprocal lattice vector whose normal points outwards the open surface. As $\bm{R}_{\rm T}$ is gauge-dependent upon the unit cell choice, we invoke $\bm{R}_{\rm L}$, namely the local polarization, that cancels the gauge dependence of $\bm{R}_{\rm T}$. Thus, we plot the number densities of topological floppy modes on the left and right open boundaries using colors of the metabeams, which are $(\nu_{\rm l},\nu_{\rm r} )=(0,2),(1,1),(2,0)$ in Figs.\ref{fig2}(d,f,g), respectively. The standard (linear) calculation of the topological polarization of the honeycomb lattice shows that the floppy modes reside only on one edge in Fig.\ref{fig2}(d) and are completely transferred to the other edge in Fig.\ref{fig2}(g) when the Guest mode is activated.

The asymmetric distribution of floppy modes governs the contrasting local stiffness on opposing boundaries. Surfaces clear of floppy modes, i.e., $\nu=0$, are as rigid as the inner bulk of the lattice, whereas boundaries that host topological floppy modes with $\nu\neq 0$ exhibit softness. Fig.\ref{fig2}(d) (Fig.\ref{fig2}(g)) shows a much softer (harder) right boundary, while Fig.\ref{fig2}(f) reflects comparable stiffness on both boundaries. Figs.\ref{fig2}(d) and (g) manifest two honeycomb lattices with identical bond orientations but opposite gear orientations (enlarged figures in~\cite{gearSI}).

When metabeams are joined together to create a hexagonal lattice, the Guest mode associated with geometric distortions of the lattice is coupled to the solitons of the individual beams in such a way that a continuous activation of the Guest mode restores the lattice to its initial configuration while reversing the polarizations of all of the beams. As a result, the global shearing Guest mode and the soliton in every metabeam together ensure $180^\circ$ gear rotations, which leads to two distinct metabeam configurations in one lattice configuration, and the complete reversal of stiffness contrast on opposing boundaries. This property is in stark contrast to previous works~\cite{vitelliPRX, PhysRevB.101.104106, PhysRevLett.117.068001, xiu2023synthetically}, where topological transitions are induced by Guest modes alone, leading to partial migration of floppy modes.

\begin{figure}[htbp]
\centering
\includegraphics[scale=0.78]{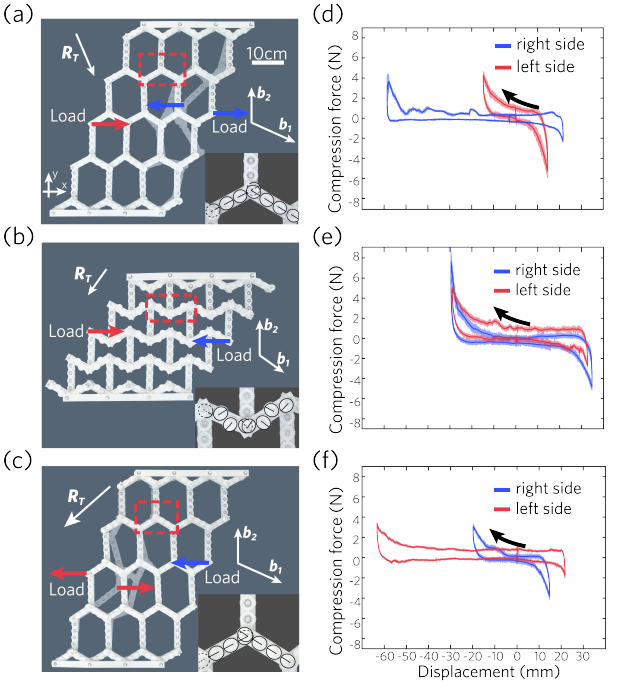}
\caption{The force-displacement measurements of the geared metamaterial in three different topological phases. (a-c) Nonlinear gear geometry and Guest mode together induce changes in $\bm{R}_{\rm T}$, with zooms depicting the geometry of the unit cell. Load arrows (blue and red) indicate push and pull tests. (d-f) Force-displacement curves with positive direction taken as $+\bm{a}_1$ (right direction) and $-\bm{a}_1$ (left direction). Colored lines show average values and shaded areas indicate standard deviation across eight measurements. 
}\label{fig3}
\end{figure}

To measure boundary stiffness, we construct the structure in Figs.\ref{fig3}(a-c) with fixed boundaries perpendicular to $\bm{b}_2$ and open boundaries perpendicular to $\bm{b}_1$, whose manual procedure is elaborated in~\cite{gearSI}. When $\bm{R}_{\rm T}$ points to the lower right, the right boundary hosts two floppy modes per supercell ($\nu_{\rm r}=2$) while the left edge shows no floppy mode ($\nu_{\rm l}=0$), in agreement with Fig.\ref{fig2}(d). Fig.\ref{fig3}(a) shows the associated metabeam configuration, the direction of the loading force, and the large deformations in gray shadow at the right boundary. In contrast, deformations on the left side are small under the same loading strength. Fig.\ref{fig3}(d) shows 8 cycles of force-displacement measurements for the structure in Fig.\ref{fig3}(a), where the deformations on the right boundary (blue curves) are much larger than that of the left edge (red curves). Using the Guest mode, which combines gears twist and bond orientation changes, we manipulate lattice topological phase transitions that stem from the nonlinear topological transition of every metabeam. In Fig.\ref{fig3}(b), both boundaries host floppy modes, corresponding to ($\nu_{\rm l}=1$, $\nu_{\rm r}=1$) in Fig.\ref{fig2}(f), and exhibit comparable stiffness as measured in Fig.\ref{fig3}(e). The Guest mode continues to transform the lattice geometry and $\bm{R}_{\rm T}$, evolving floppy modes to the final state ($\nu_{\rm l}=2$, $\nu_{\rm r}=0$) of Figs.\ref{fig3}(c,f), where the stiffness ratio between the left and right surfaces is reversed compared to Figs.\ref{fig3}(a,d). Hysteresis in the displacement (Figs.\ref{fig3}(b,d,f)) arises from gear clearance, while hysteresis in the measured force curve occurs due to friction.

\begin{figure}[htbp]
\centering
\includegraphics[scale=0.78]{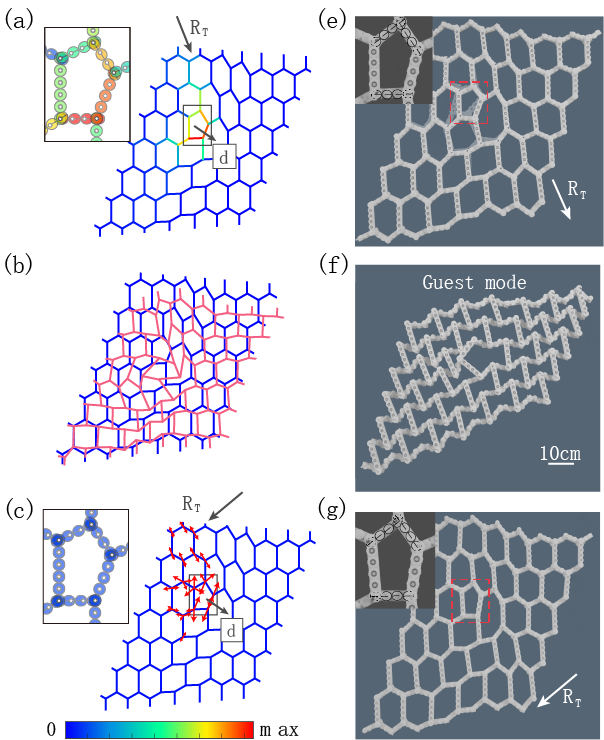}
\caption{Evolution of dislocation-bound zero modes under Guest mode. (a) Mechanical floppy mode localized around dislocation in a lattice with topological polarization $\bm{R}_{\rm T}=\bm{a}_1-\bm{a}_2$, represented by color distribution with $\bm{d}$ denoting the dislocation's dipole moment. (b) Guest mode induces changes in lattice geometry, with gear orientations in every metabeam altered accordingly. (c) State of self-stress bounded around the dislocation for a lattice with topological polarization $\bm{R}_{\rm T}=-\bm{a}_1-\bm{a}_2$, with magnitude and direction shown by red arrows. (d-f) Experimental results for (a-c).
}\label{fig4}
\end{figure}

Topologically protected mechanical zero modes can be localized around lattice dislocations. This effect stems from the interplay between two Berry phases: the dipole moment $\bm{d}$ that is perpendicular to the Burgers vector of the dislocation, and the topological polarization $\bm{R}_{\rm T}$ of the lattice. Fig.\ref{fig4} experimentally constructs a topological dislocation, around which the lattice geometry is locally modified. The number of localized mechanical zero modes around this dislocation is computed via $\nu_{\rm dislocation}=\bm{R}_{\rm T}\cdot \bm{d}/V_{\rm cell}$, where $\nu_{\rm dislocation}>0$ ($\nu_{\rm dislocation}<0$) reveals the number of mechanical floppy modes (states of self-stress), and $V_{\rm cell}$ denotes the area of the unit cell. In Fig.\ref{fig4}, the dislocation-constrained floppy modes and states of self-stress can exchange positions as the polarization $\bm{R}_{\rm T}$ is thoroughly reversed by the topological transition in the nonlinear mechanics of the metabeams.

Using other gear shapes leads to starkly different mechanics. In the triangle-trefoil-geared honeycomb metamaterial~\cite{gearSI}, the boundary stiffness is completely reversed when all gears only rotate by $\pi/3$  on the metabeams they mount on. The Guest-mode-induced lattice geometry cannot reach the auxetic state, which is in stark contrast to Fig.\ref{fig2}(f) from the elliptically-geared lattice. Therefore, gear shapes may control topological mechanical properties and manipulate other functionalities, such as negative Poisson ratio~\cite{lubensky2015phonons}.


\emph{Conclusions--}We show strongly nonlinear topological mechanics in elliptically geared metamaterials, which are ensured by quantized nonlinear Berry phases. The interplay between soliton-induced nonlinear topological mechanics and the global shearing Guest mode allows for the complete reversal of polar elasticity without disentangling the lattice. Our prototype opens up avenues for gear designs~\cite{fang2022programmable, matsumoto2022mathematical, fruchart2020dualities, frenzel2017three} with unconventional functionalities.

\emph{Acknowledgment---}This work is supported by the National Natural Science Foundation of China (Grant Nos. 12102039, 12272040).


%

\widetext

\section{Supplementary Information}

\section{Strongly Nonlinear topological mechanics of the 1D geared chain}

\subsection{Nonlinear topological index and protected floppy modes}

This section aims to derive the nonlinear topological band theory, Berry phase, and topological transitions of the elliptically geared mechanical chain, as pictorially depicted in Fig. \ref{S1} below. To this end, we first introduce a simple and abstract model called generalized nonlinear Schr\"{o}dinger equations, to establish the nonlinear topological band theory. Then we conduct the nonlinear topological mechanics of the geared metamaterial based on this band theory.

The generalized nonlinear Schr\"{o}dinger equations are constructed from a nonlinear SSH~\cite{zhou2022NC} chain composed of $N$ classical dimer fields $\Psi_n = (\Psi_n^{(1)}, \Psi_n^{(2)})$ which are nonlinearly coupled to nearest-neighbors. Under periodic boundary conditions (PBCs), the field dynamics are governed by the following equations of motion
\begin{eqnarray}\label{A1}
\mathrm{i}\partial_t \Psi_n^{(1)}  & = &  f_e(\Psi_n^{(2)})-f_{-e}(\Psi_{n+1}^{(2)}),\nonumber \\
\mathrm{i}\partial_t \Psi_n^{(2)}  & = &  f_e(\Psi_n^{(1)})-f_{-e}(\Psi_{n-1}^{(1)}),
\end{eqnarray}
where the interaction $f_e(\Psi) = s_e({\rm Re\,}\Psi)+\mathrm{i}\,s_e({\rm Im\,}\Psi)$ is a nonlinear function of the complex field variable $\Psi$, the function
\begin{eqnarray}\label{A2}
s_e(x) = a(1-e^2) \int_0^x dx' \frac{(1+2e\cos x'+e^2)^{1/2}}{(1+e\cos x')^2}
\end{eqnarray}
stems from the arc length of an ellipse that will appear later in the consideration of the gear mechanics, $a$ is the semi major-axis of the elliptic gears, and $e$ is the eccentricity that yields $|e|<1$.

The topological properties of Eqs. (\ref{A1}) are studied as follows. The nonlinear bulk modes are spatial-temporal periodic. They take the traveling plane-wave form~\cite{zhou2022NC}, 
\begin{eqnarray}\label{A3}
\Psi_q = (\Psi_q^{(1)}(\omega t-qn), \Psi_q^{(2)}(\omega t-qn+\phi_q)),
\end{eqnarray}
where $\omega$ and $q$ are the frequency and wavenumber, respectively. $\Psi_q^{(j=1,2)}(\theta)$ are $2\pi$-periodic wave components, where the phase conditions are chosen by asking ${\rm Re\,}\Psi_{q}^{(j)}(\theta=0)=A$, and $A\overset{\rm def}{=}\max({\rm Re\,}\Psi_q^{(j)})$ defines the mode amplitude. Following this condition, $\phi_q$ characterizes the relative phase between the two wave components. Nonlinear bulk modes are not sinusoidal functions of time, which is a natural result of strong nonlinearities. Following the mathematical formalism of Ref.~\cite{zhou2022NC}, we realize the adiabatic evolution of the nonlinear bulk modes by asking the wavenumber to traverse through the Brillouin zone. Consequently, the geometric phase acquired by the process of adiabatic evolution, namely nonlinear Berry phase, is presented as follows,
\begin{eqnarray}\label{A4}
\gamma(A) & = & 
\oint_{\rm BZ}dq \frac{\sum_{l\in\mathcal{Z}} \left( l |\psi_{l,q}^{(2)} |^2 \frac{\partial\phi_q}{\partial q}+\mathrm{i}\sum_{j}\psi_{l,q}^{(j)*}\frac{\partial\psi_{l,q}^{(j)}}{\partial q}\right)}
{\sum_{l'\in\mathcal{Z}}  l'  \left(\sum_{j'}|\psi_{l',q}^{(j')}|^2\right)},\nonumber \\
& = & 
\pi\Theta(s_e(A)-s_{-e}(A))\nonumber \\
& = & 
\left\{
\begin{aligned}
\pi\,[1+{\rm sgn}(e)]/2 & {} & A<A_c \\ 
{\rm ill-defined} & {} & A=A_c\\
\pi\,[1-{\rm sgn}(e)]/2 & {} & A>A_c \\ 
\end{aligned}
\right.,
\end{eqnarray}
where $j, j'=1,2$ denote the two wave components, $\psi_{l,q}^{(j)} = (2\pi)^{-1}\int_0^{2\pi} e^{\mathrm{i} l\theta} \Psi_q^{(j)}d\theta$ is the $l$-th Fourier component of $\Psi_q^{(j)}$, and $\Theta(x)$ is the step function. ${\rm sgn}(e)$ represents the sign of the eccentricity of the ellipse. Specifically, $e>0$ and $e<0$ correspond to the right-sided and left-sided focal points, respectively. Nonlinear Berry phase $\gamma(A)$ is quantized to be $0$ or $\pi$, and is controlled by the amplitude $A$ of nonlinear bulk modes. The topological transition amplitude $A=A_c$ occurs at the critical point when $\gamma$ becomes ill-defined, which in turn demands $s_{-e}(A_c) = s_e(A_c)$ to have $A_c=\pi$. This critical amplitude is the natural result of the mirror symmetry in elliptic gears. 


In the linear regime when $A\ll A_c$, $s_e(\Psi)\approx a(1-e) \Psi$. One can convert the equations of motion from real space to momentum space, to reach the Hamiltonian $H_q = h_q\sigma_++h_{-q}\sigma_-$, where $\sigma_\pm = (\sigma_x\pm\mathrm{i}\sigma_y)/2$, $\sigma_{x,y,z}$ are Pauli matrices, and $h_q = a(1-e)-a(1+e)e^{\mathrm{i}q}$. The topological invariant is reduced to the conventional form  $\gamma(A\ll A_c)=\pi\Theta (e)=\frac{\mathrm{i}}{2}\oint_{\rm BZ} \mathrm{d}q \, \partial_q \ln \det h_q=\pi$, which is perfectly in agreement with the winding number derived in isostatic lattices.

As the amplitudes of nonlinear bulk modes grow, this topological invariant $\gamma(A<A_c)=\pi$ cannot change until the amplitudes reaches the transition point $A=A_c$. 
Above $A_c$, nonlinear Berry phase is well-defined again to take the trivial value $\gamma(A>A_c)=0$. According to the nonlinear extension of bulk-boundary correspondence, the localization of boundary modes are guaranteed by the topological invariant derived from the bulk bands. This correspondence is directly confirmed here by analytically finding the boundary solution under open boundary conditions (OBCs). In agreement with the invariant $\gamma(A<A_c)=\pi$, nonlinear topological floppy modes emerge on the left open boundary for $|{\rm Re\,}\Psi_n^{(1)}|$ and $|{\rm Im\,}\Psi_n^{(1)}|<A_c$. As indicated by the topological number $\gamma(A>A_c)=0$, nonlinear boundary mode cannot emerge for either $|{\rm Re\,}\Psi_n^{(1)}|>A_c$ or $|{\rm Im\,}\Psi_n^{(1)}|>A_c$. 

\renewcommand{\thefigure}{S1}
\begin{figure}[htbp]
\centering
\includegraphics[scale=0.25]{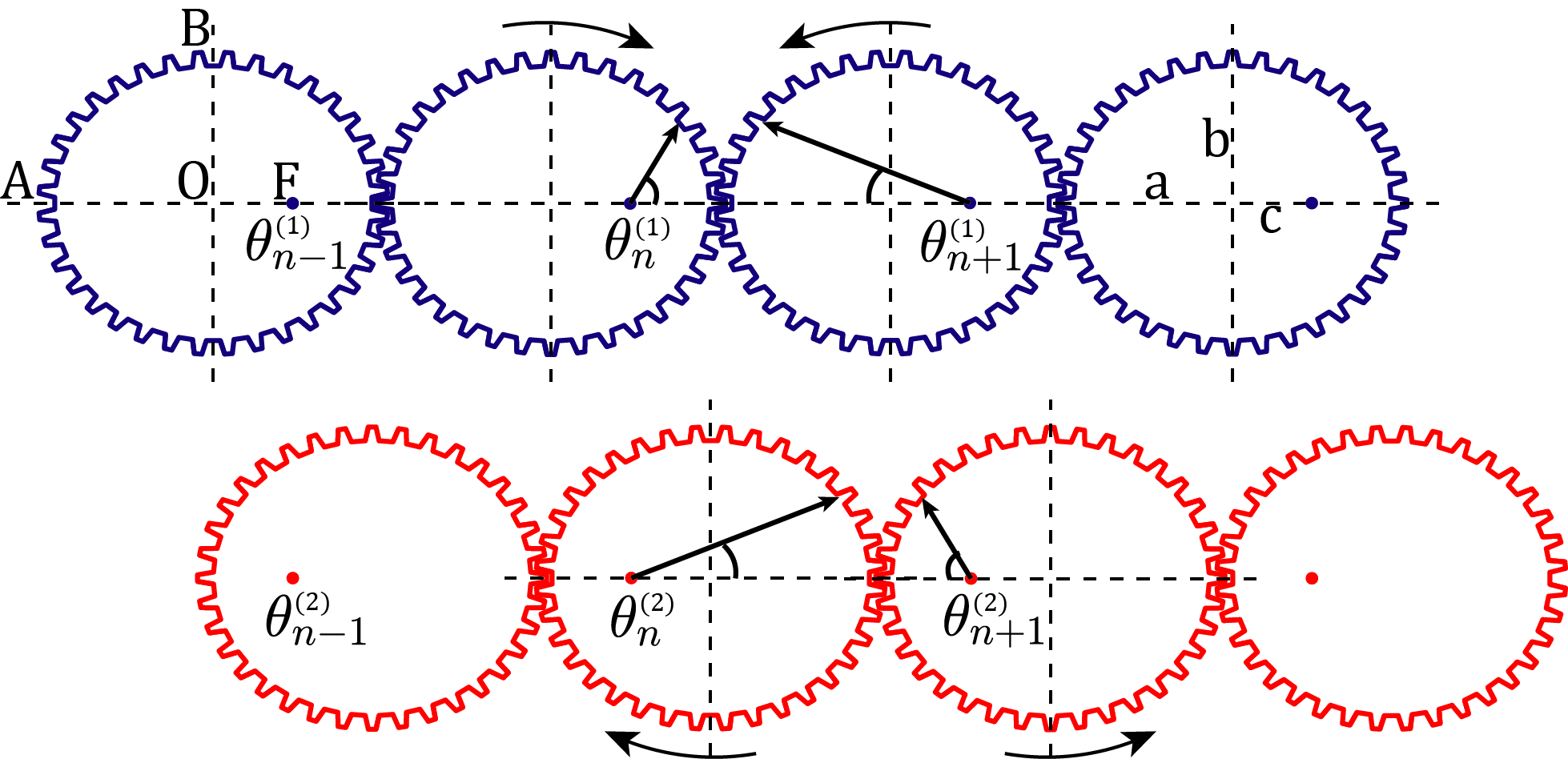}
\caption{Elliptically geared mechanical chains that feature nonlinear topological evanescent floppy modes. The modes are localized on the left and right open boundaries for the blue and red chains, respectively.
}\label{S1}
\end{figure}




We now employ the nonlinear topological band theory derived from the generalized nonlinear Schr\"{o}dinger equations, to uncover the topological properties of the geared mechanical lattice. The system consists of two geared mechanical chains subjected to OBCs on both ends. Blue and red chains are decoupled to one another. The unit cell is composed of two elliptic gears with the eccentricity $e$. Blue and red gears are freely rotatable about fixed pivots on the right-sided and left-sided focal points, respectively. O denotes the center, F denotes the focal point, and $\overline{\rm OA}=a$, $\overline{\rm OB}=a\sqrt{1-e^2}$, $\overline{\rm OF}=ae$ are the semi-major axis, semi-minor axis, and focal distance of the ellipse, respectively. 
To bridge Eqs. (\ref{A1}) and the mechanics of elliptically geared chains, we notice that Eqs. (\ref{A1}) can be decomposed to the following form, 
\begin{eqnarray}\label{A5}
-\partial_t \alpha_n & = & s_e(\beta_n)-s_{-e}(\beta_{n+1}),\nonumber \\
\partial_t \beta_n & = & s_e(\alpha_n)-s_{-e}(\alpha_{n-1}),
\end{eqnarray}
where the field variables 
\begin{eqnarray}\label{A6}
(\alpha_n, \beta_n) = (-{\rm Re\,}\Psi_n^{(1)}, {\rm Im\,}\Psi_n^{(2)}),\,\,\, {\rm and}\,\,\, ({\rm Im\,}\Psi_n^{(1)}, {\rm Re\,}\Psi_n^{(2)})
\end{eqnarray}
in Eqs. (\ref{A5}), such that Eqs. (\ref{A5}) in fact manifest a total of four equations that restore Eqs. (\ref{A1}) above. On the other hand, 
the sliding distances of the engaging gears are denoted as $\ell_n^{(1)}$ and $\ell_n^{(2)}$ for blue and red gears, respectively. They are expressed as 
\begin{eqnarray}\label{A7}
\ell_n^{(1)} & = & s_e(\theta_n^{(1)})-s_{-e}(\theta_{n+1}^{(1)}),\nonumber \\
-\ell_{n-1}^{(2)} & = & s_e(\theta_{n}^{(2)})-s_{-e}(\theta_{n-1}^{(2)}), 
\end{eqnarray}
where the rotation angles are defined in an alternative way $\theta_n\to (-1)^n\theta_n$ in order to simplify the mathematics, 
and $s_e(\theta)$, $s_{-e}(\theta)$ are the arc lengths for the rotation angle $\theta$ to the right-sided and left-sided focal points of the coupling gears, respectively. In the linear regime when $\theta\ll \pi$, we have $s_e(\theta) \approx a(1-e) \theta$. Thus Eqs. (\ref{A7}) reduce to the linear form,
\begin{eqnarray}\label{A8}
\ell_n^{(1)} & = & a(1-e) \theta_n^{(1)}-a(1+e) \theta_{n+1}^{(1)},\nonumber \\
 - \ell_{n-1}^{(2)}  & = & a(1-e) \theta_{n}^{(2)}-a(1+e) \theta_{n-1}^{(2)}. 
\end{eqnarray}
We now define the mechanical impedance which is analogous to the impedance of electrical circuits~\cite{liu2020octupole}, $Z_n^{(j)} = \ell_{n}^{(j)}/\theta_n^{(j)}$ for $j = 1,2$. It quantitatively describes the relationships between sliding distances and rotation angles of the gears. In the linear regime, $Z_n^{(j)}$ is analogous to the Hamiltonian of the linearized model in Eqs. (\ref{A1}), which offers quantized Berry phase and linear topological boundary floppy modes~\cite{kane2014topological,vitelliPRX}. To elucidate this, we consider the linearized mechanical impedance in momentum space, $Z(q) = a(1-e) +a(1+e) e^{\mathrm{i}q}$, which grants quantized Berry phase $\gamma=\frac{\mathrm{i}}{2}\oint_{\rm BZ} \mathrm{d}q \, \partial_q \ln \det Z(q)=0$ for $e<0$ or $\pi$ for $e>0$. The topological attributes and floppy mode are predicted by the quantized Berry phase. In particular, $\gamma=\pi$ and the mechanical floppy mode is exponentially localized on the left boundary of the blue chain, whereas for $\gamma=0$, the left boundary of the blue chain is free of mechanical floppy modes. In addition, topological transition occurs when the phonon band gap is closed by a floppy bulk mode.  
In conclusion, both of the linearized Hamiltonian in Eqs. (\ref{A1}) and the linear mechanical impedance capture the topological index and the localization of zero-frequency evanescent modes. 

Since the linearized mechanical impedance correctly predicts the topological floppy modes, it is natural to extend this correspondence to the nonlinear regime. According to Eq. (\ref{A4}), there is a critical amplitude $A_c = \pi$, below which nonlinear Berry phase takes the non-trivial integer value. Likewise, nonlinear topological floppy modes that enable $\ell_n=0$ for all $n$ in Eqs. (\ref{A7}) can emerge as long as the rotation angle of the first gear yields $\theta_1<A_c=\pi$. At $\theta_1^{(1)}=A_c=\pi$, a bulk floppy mode, namely $\theta_1=A_c=\pi$, closes the nonlinear band gap of the geared isostatic chain. $\gamma(A_c)$ is ill-defined by the degeneracy in Brillouin zone at zero frequency. 
Finally, as stated by the trivially-valued index $\gamma(A>A_c)=0$, the left open boundary of the blue chain is free of localized floppy modes for $\theta_1^{(1)}>A_c$. Meanwhile, the red chain is rigid on its left boundary and therefore mechanical floppy modes are non-excitable, leading to $\theta_n^{(2)}\equiv 0$ for all cases. The spatial profiles of $\theta_n^{(j=1,2)}$ derived from Eqs. (\ref{A7}) agree perfectly well with the nonlinear topological modes derived from Eqs. (\ref{A5}). In summary, geared nonlinear isostatic chains proposed in Fig. \ref{S1} host nonlinear topological floppy modes, which are guaranteed by the nonlinear topological invariant derived from the generalized nonlinear Schr\"{o}dinger equations.





\subsection{Topological floppy modes insensitive to disorders}

\renewcommand{\thefigure}{S2}
\begin{figure}[htb]
\centering
\includegraphics[scale=0.8]{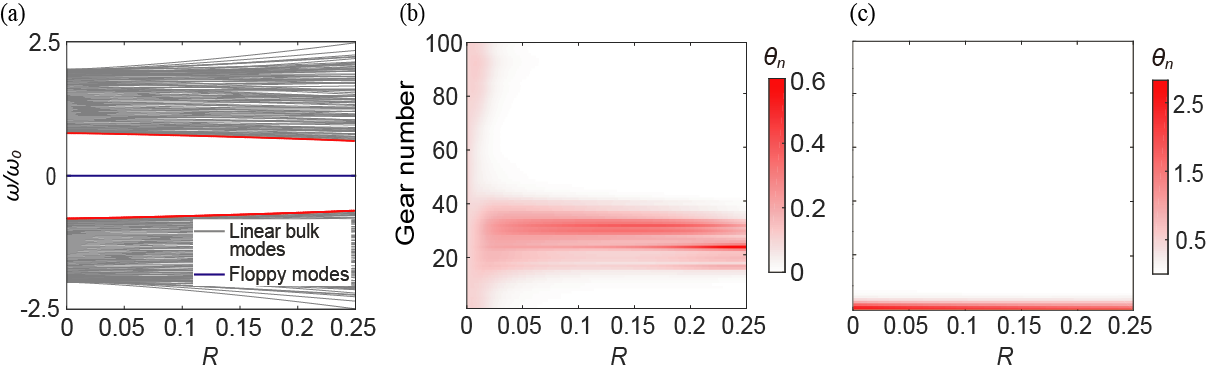}
\caption{The evolution of bulk states and topological boundary modes in the presence of disorder. (a) The frequencies of bulk modes and floppy mode are plotted in grey curves and the blue line, respectively. Two red curves mark the eigenfrequencies of lowest-frequency bulk modes. (b) The spatial profile of the linear bulk state evolves with randomness. (c) The strongly nonlinear floppy mode profile remains unchanged as randomness increases. 
}\label{S2}
\end{figure}

The topological invariants derived from the nonlinear bulk bands are guaranteed to be integer multiples of $\pi$. 
Consequently, topological boundary modes are robust against system disorders. To demonstrate the robustness, we numerically study a chain of $N=100$ elliptic gears subjected to OBCs. We introduce disturbances in the compatibility matrix. For different site number $n$, the random gear parameters, such as the semi major-axis $a_n$ and the eccentricity $e_n$, fluctuate around their average values $\langle a\rangle$ and $\langle e\rangle$ via $a_n=(1+r_n )\langle a\rangle$ and $e_n=(1+r_n )\langle e\rangle$. Here, $r_n$ is a random number for different sites with $|r_n|\le R$, and the average values for the $a$ and $e$ are $\langle a\rangle = 1$ and $\langle e \rangle = 0.4$. For small rotation angles $\theta_n\approx 0$, the eigenfrequencies of linear normal modes are numerically computed by increasing the randomness $R$ from 0 to 25$\%$. Fig. S2(a) depicts the band structure of the ``square-root of dynamical matrix"~\cite{kane2014topological}
\begin{eqnarray}\label{A20}
H = \left(\begin{array}{cc}0 & \bm{C}^\top\\ \bm{C} & 0\end{array}\right),
\end{eqnarray}
where $\bm{C}$ is the real-space representation of the compatibility matrix of the underlying random chain. It is defined by relating the sliding distances and gear rotations via $\ell_{n_1} = \bm{C}_{n_1,n_2}\theta_{n_2}$. From Eqs. \ref{A8}, the matrix elements are itemized as $\bm{C}_{n_1,n_2} = a_{n_2} (1-e_{n_2})\delta_{n_1,n_2}-a_{n_2}(1+e_{n_2})\delta_{n_1+1,n_2}$. 

$H$ is subjected to an intrinsic chiral symmetry~\cite{kane2014topological}, which guarantees eigenstates to come in $\pm\omega/\omega_0$ pairs and the frequency of topological modes be zero (here, we denote $\omega_0 =\sqrt{ka^2/I}$ as the frequency unit). We can see in Fig. \ref{S2}(a) that as the randomness increases, the frequencies of linear bulk modes (grey curves) deviate from the initial values while the frequency of the linear topological floppy mode (blue line) remains zero. Fig. \ref{S2}(b) shows that the spatial profile of the lowest linear bulk states (two red curves in Fig. \ref{S2}(a)) changes dramatically with the magnitude of randomness. In contrast, as shown in Fig. \ref{S2}(c), even $\theta_1$ increases up to 2.8 (the degree of nonlinearity $\epsilon=0.52$), the nonlinear topological floppy mode can still be excited upon the adiabatic rotation of the first gear, and the location of the mode is immune to this disturbance. These two phenomena demonstrate the robustness of the strongly nonlinear topological mode.

\subsection{Soliton mode in the 1D geared chain}

\renewcommand{\thefigure}{S3}
\begin{figure}[htb]
\centering
\includegraphics[scale=0.7]{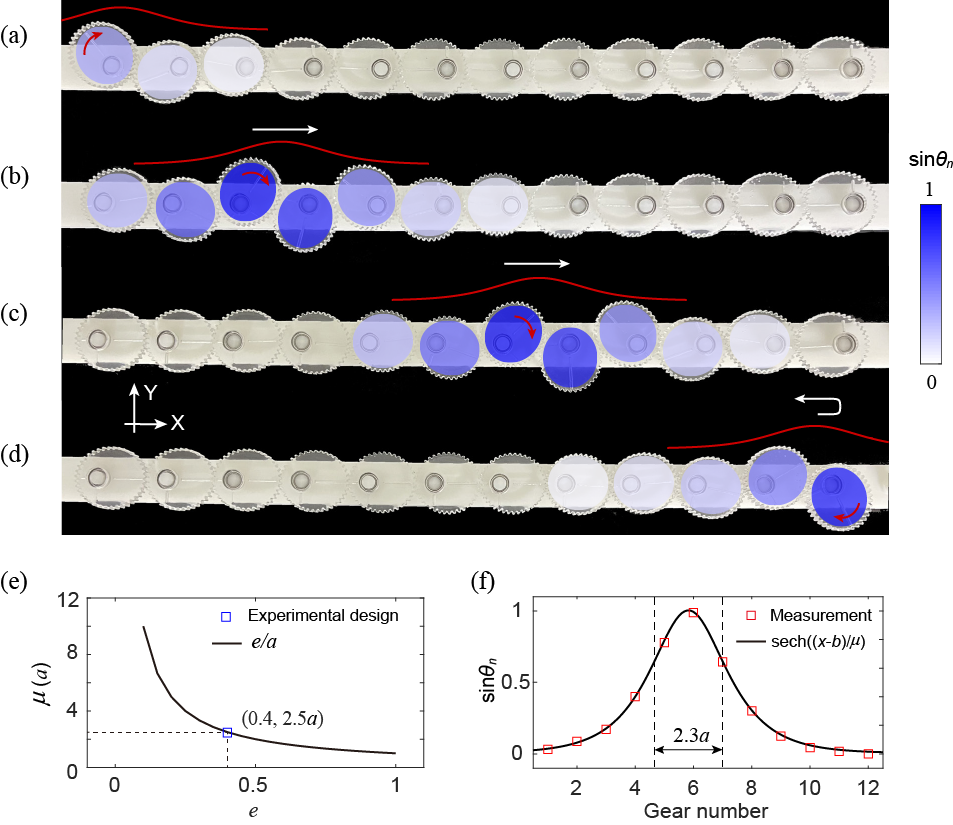}
\caption{The soliton. (a)-(d) The rotations of the gears constitute a soliton in the chain. As $\theta_1$ approaches to $\pi$, the soliton propagates from the left end to the right side of the chain. (e) The width of soliton versus the eccentricity of the elliptic gears. The blue square mark corresponds to the eccentricity in our experimental fabrication. (f) The spatial distribution of the soliton along the chain. The red square marks show the experimental measurements and the black curve corresponds to the fitting function of ${\rm sech}((x-b)/\mu)$.}\label{S3}
\end{figure}

The main text addresses the propagation of a soliton from the experimental perspective. Here we demonstrate the soliton propagation from the theoretical aspect. To this end, we introduce a number of useful approximations. First, we consider the small-eccentricity limit with $e\ll 1$ to approximate the arc length as follows, 
\begin{eqnarray}
s_e(\theta)
 \approx 
a (\theta-e\sin \theta).
\end{eqnarray}
We re-define the orientations of the gears relative to the upward or downward normals, via $\Delta\theta_n = \theta_n-\pi/2$. We consider the $\Delta\theta_n\ll \pi$ limit, to further approximate the arc length as 
\begin{eqnarray}
s_e(\theta) \approx a (\pi/2-e\cos\Delta\theta+\Delta\theta).
\end{eqnarray}
Next, we adopt the long-wavelength approximation 
\begin{eqnarray}
\Delta\theta_{n+1}-\Delta\theta_n \approx 2a\partial_x\Delta\theta(x)
\end{eqnarray}
to reduce the sliding distance as follows,
\begin{eqnarray}
s_e(\theta_n)-s_{-e}(\theta_{n+1}) \approx 
-2a (a\partial_x\Delta\theta+e\cos\Delta\theta).
\end{eqnarray}
Equipped with all the above-mentioned approximations, the system Lagrangian now reads 
\begin{eqnarray}
L & = & \sum_{n=1}^N \left[\frac{1}{2}I\dot{\theta_n}^2- V(s_e(\theta_n)-s_{-e}(\theta_{n+1}))\right] \nonumber \\
 & \approx & 
\int_0^{2Na}\frac{dx}{2a}\left[ \frac{1}{2}I\Delta\dot{\theta}^2- V(2a (a\partial_x\Delta\theta+e\cos\Delta\theta)) \right]\nonumber \\
 & \approx & 
\int_0^{2Na}\frac{dx}{2a}\left[ \frac{1}{2}I\Delta\dot{\theta}^2- 2ka^4\left( \partial_x\Delta\theta+\frac{e}{a}\cos\Delta\theta\right)^2 \right],
\end{eqnarray}
with the Lagrange equations of motion 
\begin{eqnarray}\label{solitonEOM}
I\partial_t^2\Delta\theta-4ka^4\left(\partial_x^2\Delta\theta+\frac{e^2}{a^2}\sin\Delta\theta\cos\Delta\theta\right)=0.
\end{eqnarray}
The standard kink soliton solution of Eq.(\ref{solitonEOM}) is given by 
\begin{eqnarray}
\sin\Delta\theta(x) =- \tanh\left(\frac{x-vt}{\mu \sqrt{1-v^2/c^2}}\right),
\end{eqnarray}
where $c = \sqrt{4ka^4/I}$ is the speed of light, and $\mu = a/e$ is the width of the soliton. 
Using $\theta_n = \Delta\theta_n+\pi/2$, it is straightforward to show  
\begin{eqnarray}
\sin\theta(x) = {\rm sech}\left(\frac{x-vt}{\mu\sqrt{1-v^2/c^2}}\right).
\end{eqnarray}
This result indicates that for gears far away from the soliton center, the left-sided gear lean rightwards ($\sin\Delta\theta(x\to-\infty)=+1$ and $\theta(x\to-\infty) = \pi$), and the right-sided gears lean leftwards ($\sin\Delta\theta(x\to+\infty)=-1$ and $\theta(x\to+\infty)=0$). We find that this analytical prediction of the soliton profile is in agreement with the experimental measurement obtained from our elliptically geared chain. Fig. \ref{S3}(d) presents the end state of this soliton propagation.

\subsection{Experimental setup for torque measurements}

In order to measure the torque at the open boundary, we apply a digital torque to the gear at the open boundary, and attach a clockwork spring to the opposite boundary. We read the digital torque each time the number of gears in the chain is increased. The measurements of two different boundary stiffness, namely the rigid and floppy ones, are presented in Figs.\ref{S4}(a) and (b) respectively.

\renewcommand{\thefigure}{S4}
\begin{figure}[htb]
\centering
\includegraphics[scale=0.75]{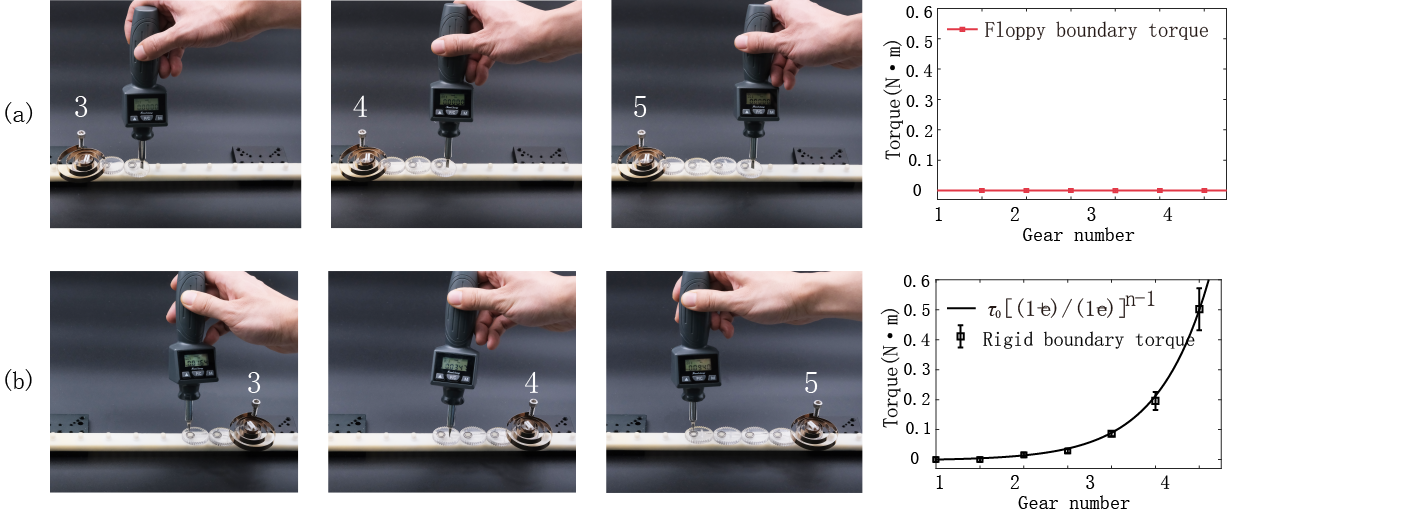}
\caption{The setup for the floppy torque measurement and rigid torque measurement.
}\label{S4}
\end{figure}

In Fig. \ref{S4}(a), a clockwork spring is attached to the left rigid boundary to measure the local stiffness of the floppy boundary on the right side. Then, the gear at the right end of the chain is rotated by digital wrench, with each increase in gear number being recorded along with the externally applied torque. Despite the increase of the chain's gear number, the torques applied to the floppy boundary on the right remain at zero. In Fig. \ref{S4}(b), a clockwork spring is attached to the right floppy boundary to measure the local stiffness of the rigid boundary on the left side. The applied torques, as indicated by the black dots, exhibit a significant increase with the number of the gears.

The exponential growth of stiffness on the rigid boundary as a function of gear number can be mathematically modeled as follows. Let us consider an elliptical gear with eccentricity $e<1$. The torque-balance condition on the gear can be written as $F_{\rm l} r_{\rm l} = F_{\rm r} r_{\rm r}$, where $F_{\rm l}$ and $F_{\rm r}$ are the applied forces on the left and right sides of the gear, respectively, and $r_{\rm l}$ and $r_{\rm r}$ are the corresponding lever arms. For an elliptical gear, $r_{\rm l}$ and $r_{\rm r}$ are given by $r_{\rm l} = a(1+e)$ and $r_{\rm r}=a(1-e)$, where $a$ is the semi-major axis of the gear. We consider a chain of $N$ elliptical gears and apply an initial torque $M_1$ to the first gear. Since each gear is torque-balanced, the left and right tangential forces exerted on them are related by $F_{\rm r}/F_{\rm l}=(1+e)/(1-e)$, leading to a transmission rate of torques along the chain given by $M_{n+1}/M_n = (1+e)/(1-e)$. Therefore, in order to maintain torque balance in the $N$th gear, a torque $M_N = M_1 [(1+e)/(1-e)]^{N-1}$ is required to be applied to the last gear. This expression describes the relationship between the applied torques at each gear position.

\subsection{Topological phase transition amplitudes of triangle-trefoil geared chains}

As shown in Fig. \ref{S5}, the chain consists of conjugated gears of triangles and trefoils, which rotate freely about their fixed pivots. During the rotation, the adjacent triangle and trefoil have zero sliding. At first, the floppy mode is located at left side of the chain (shown in Fig. \ref{S5}(a)). As the rotation angle of the first trefoil gear grows, the gear rotations penetrate into the chain. When the rotation angle of the first trefoil approaches $A_c=\pi/3$, the floppy mode transits into the nonlinear bulk mode (Fig. \ref{S5}(d)). We note that the critical transition amplitude $A_c=\pi/3$ originates from the $C_3$ symmetry in the triangle and trefoil gears. As such, the critical transition amplitude $A_c$ can take values other than $\pi$. 

\renewcommand{\thefigure}{S5}
\begin{figure}[htb]
\centering
\includegraphics[scale=0.73]{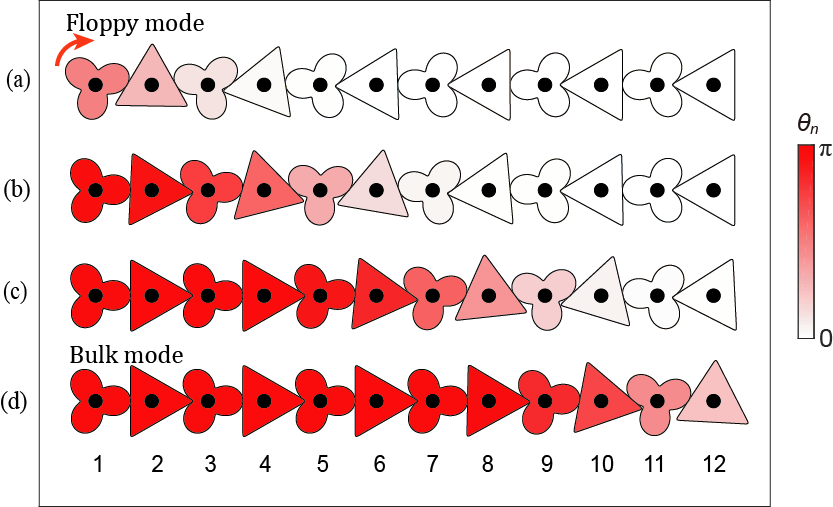}
\caption{Topological phase transition of triangle-trefoil geared chain. As the rotation angle of the first trefoil gear grows from (a) to (d), the boundary mode gradually penetrates into the chain. In section (d), the nonlinear bulk mode runs through the whole chain and the corresponding nonlinear band gap closes.
}\label{S5}
\end{figure}

\renewcommand{\thefigure}{S6}
\begin{figure}[htb]
\centering
\includegraphics[scale=0.8]{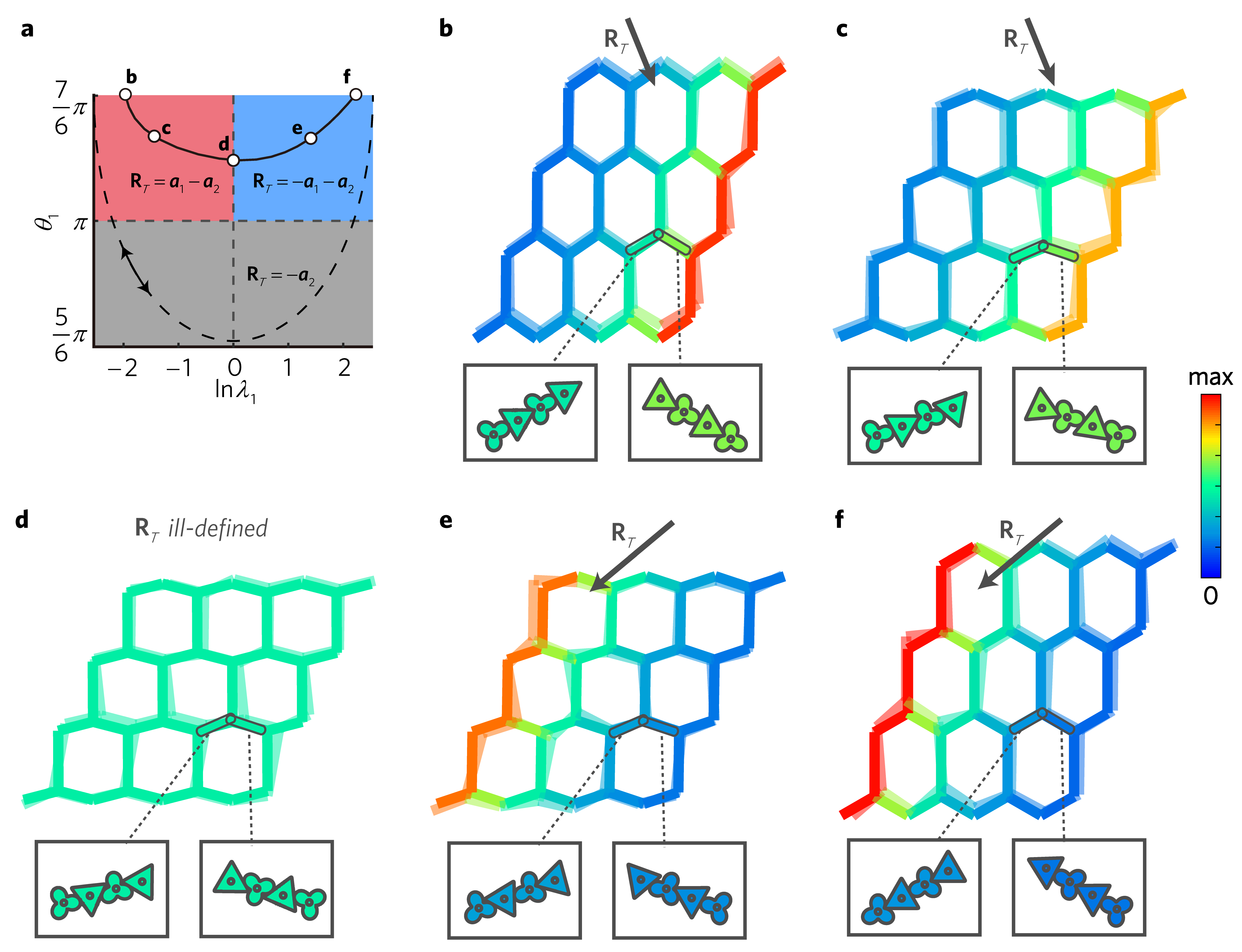}
\caption{Topological phases of the 2D honeycomb lattice that is constructed from the triangle-trefoil-geared metabeams. (a) The multi-phase diagram of the topological mechanics, under the parameter constraints of $\theta_2=\pi-\theta_1$ and $\lambda_2=\lambda_1^{-1}$. We use the topological phase diagram from the elliptically-geared metamaterial in the main text (i.e., red, blue, and grey-shaded areas) to compare with the topological phases of the triangle-trefoil-geared metamaterials. The black dashed line represents how the topological polarization of the elliptically-geared system changes under the shearing Guest mode, whereas the black solid line depicts that of the triangle-trefoil-geared lattice. White dots correspond to lattice configurations in panels (b-f). (b-f), Geometric configurations of the triangle-trefoil-geared metamaterial, where the topological polarizations are $\bm{R}_{\rm T}=\bm{a}_1-\bm{a}_2$ in (b) and (c), ill-defined in (d), and $\bm{R}_{\rm T}=-\bm{a}_1-\bm{a}_2$ in (e) and (f), respectively. The insets depict the corresponding metabeam configurations.}\label{S5v2}
\end{figure}

We use the triangle-trefoil-geared metabeams to assemble the honeycomb lattice. The beam orientations remain the same as those in Fig.2 of the main text. To be specific, the initial lattice configuration, as shown in Fig. \ref{S5v2}(a), has the bond orientations $\theta_1=7\pi/6$, $\theta_2=-\pi/6$, and $\theta_3=\pi/2$. Type 1 and 2 metabeams are prepared by assembling $N_1=2$ and $N_2=2$ \emph{pairs of} triangle-trefoil gears (i.e., we assemble $N_1=N_2=2$ triangle and trefoil gears on metabeams type 1 and 2). We assemble $N_3=6$ circular gears on metabeam type 3. As a result, the transmission rates in Fig. \ref{S5v2}(a) are $\lambda_1=1/\lambda_2=9.04$ and $\lambda_3=1$.

We theoretically study the triangle-trefoil-geared honeycomb lattice. The lattice is on the verge of mechanical instability as it has the balanced degrees of freedom and constraints. We numerically evaluate the traveling arc length of the contacting point between adjacent gears, in terms of the gears’ angular displacements. This methodology carries out the nonlinear relationship among the compatibility matrix, the shearing Guest mode angle, and the degree of nonlinearity. This nonlinear relationship is analogous to Eq. \ref{guest_4}, which is derived from the elliptically geared metabeam. The only difference is that the elliptic traveling arc length, $\Theta(x)$ in Eq. \ref{guest_2}, is now replaced by the traveling arc length of the contacting point between adjacent gears along the triangular gear.
 
Using the Guest-mode-controlled compatibility matrix, we compute the topological polarization $\bm{R}_{\rm T}$ of the triangle-trefoil-geared 2D lattice, which follows the black solid trajectory in the topological mechanical phase diagram of Fig. \ref{S5v2}(a).

As indicated by Fig. \ref{S5v2}(a), the topological phase of the elliptically-geared lattice follows the black dashed trajectory in the parameter space, whereas the triangle-trefoil-geared metamaterial follows the black solid one, leading to starkly distinct mechanical properties. As shown by Fig. 2(f) of the main text, the elliptically-geared honeycomb structure can reach an \emph{auxetic} state, whose topological polarization, $\bm{R}_{\rm T}=-\bm{a}_2$, indicates that floppy modes emerge on both of the parallel open boundaries with comparable edge stiffness. In contrast, the triangle-trefoil-geared metamaterial cannot reach such an auxetic state (Figs. \ref{S5v2}(c-e)), and the floppy modes always arise exclusively at a single boundary. These results indicate significantly distinct mechanical responses in metamaterials using other gear shapes, such as the negative Poisson ratio and polarized boundary elasticity. 

The fundamental reason behind such mechanical distinction lies in the relatively smaller topological transition amplitude, $A_c=\pi/3$, in the triangle-trefoil metabeams, comparing to elliptically-geared metabeams with $A_c=\pi$. During the procedure of Guest-mode-shearing, the change of beam orientations in the triangle-trefoil-geared structure is smaller ($\theta_1$ remains above $180^\circ$) than that of the elliptically-geared metamaterial ($\theta_1$ drops below $180^\circ$). Consequently, the triangle-trefoil-geared structure cannot achieve the auxetic state. 

The topological polarization of the lattice is governed by both beam orientations and transmission rates. However, when the transmission rate is close to one (i.e., when $\ln\lambda_1$ approaches 0), the topological polarization is dominated by beam orientations only. The topological polarization of the elliptically-geared metamaterial can reach $\bm{R}_{\rm T}=-\bm{a}_2$, because the change of Guest mode angle is large and can drop below $180^\circ$. In contrast, the polarization in triangle-trefoil-geared metamaterial cannot reach this phase, as the change of the Guest mode angle is small and remains above $180^\circ$.


\section{Elliptically geared topological metamaterials in two-dimensions}

\subsection{Compatibility matrix, Guest modes, and the topology of lattice mechanics}

Based on the nonlinear topological mechanics of the 1D geared chain, we take advantage of this unique platform to design novel topological mechanical metamaterials in two-dimensions and reveal novel physics. Below we illustrate the topological boundary floppy modes using generalized honeycomb lattice, as pictorially depicted in Fig. \ref{S6}. 

\renewcommand{\thefigure}{S7}
\begin{figure}[htb]
\centering
\includegraphics[scale=0.4]{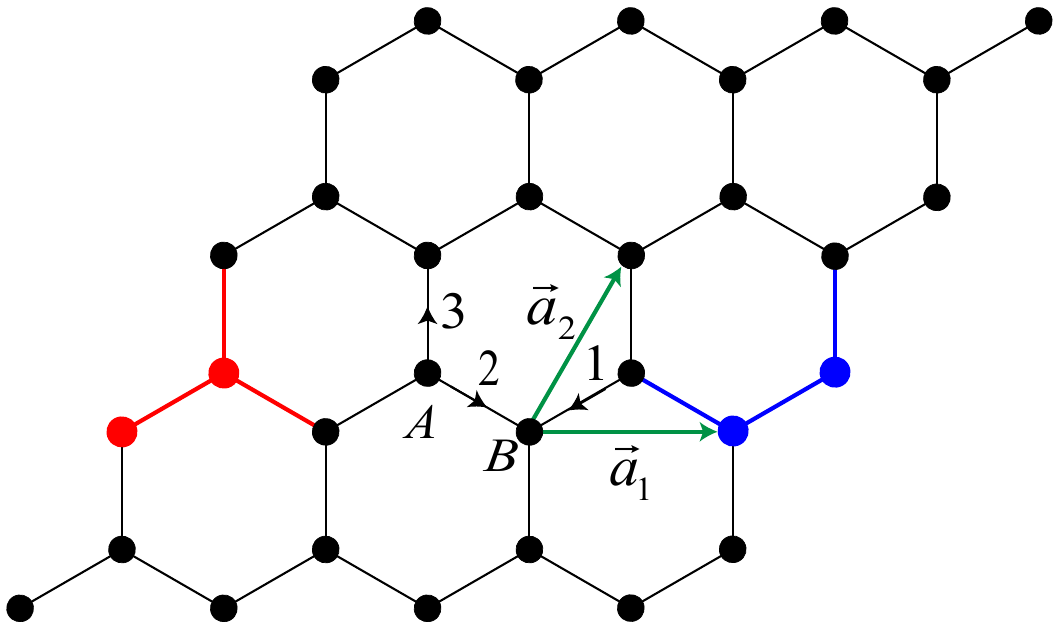}
\caption{Schematic illustration of the two-dimensional elliptically geared lattice, where OBC is adopted in $\bm{a}_1$ and PBC is in $\bm{a}_2$. The unit cell contains two sites, denoted as $A$ and $B$, and three metabeams labelled by 1, 2, and 3. Black arrows stand for the unit vectors of beam orientations. The primitive vectors $\bm{a}_1$ and $\bm{a}_2$ are marked by green arrows. 
The unit cells that are compatible with the left and right open boundaries are marked by red and blue, respectively, which defines the local polarization vector $\bm{R}_{\rm L}$. 
}\label{S6}
\end{figure}

To discuss the collective motion of lattice mechanics, we conduct a single metabeam depicted by Fig. \ref{S7} that is freely movable in 2D space. Thus, the beam is freely rotatable with the orientation denoted as $\hat{n} = (\cos\theta,\sin\theta)$. The transverse direction of the beam is naturally denoted as $\hat{t} = (-\sin\theta, \cos\theta)$. The two ends of the beam are denoted as sites $a$ and $b$, from which we build the relationship between site movements and beam elasticity. Since the beam is anchored by a chain of gears, each end hosts three degrees of freedom, namely two translational modes $\bm{u}_a = (u_{ax}, u_{ay})$ ($\bm{u}_b = (u_{bx}, u_{by})$), and one rotational degree of freedom denoted as $u_{a\theta}$ ($u_{b\theta}$). We ask $|\bm{u}_{a,b}|\ll L$ 
and $u_{a,b \, \theta}\ll \pi$ to enable linear elastic theory. On the other hand, every metabeam provides a longitudinal and a transverse constraint. The longitudinal constraint indicates that the node displacements $\bm{u}_a$, $\bm{u}_b$ and the beam elongation $e_n$ are not arbitrary. They are related by $e_n = (\bm{u}_b-\bm{u}_a)\cdot\hat{n}$. The transverse constraint is carried out as follows. If the beam is not allowed to rotate, the rotations of $A$ and $B$ gears yield the constraint $\lambda u_{a\theta}+u_{b\theta}=e_t$, where the small-rotation approximation has been adopted from Eqs. (\ref{A7}), $e_t$ denotes the sliding distance between the gears, and $\lambda$ is the transmission ratio determined by the gear orientations. For example, if all gears are aligned in the orientation of the metabeam, the transmission ratio is $\lambda=((1-e)/(1+e))^{n-1}$, where $n$ denotes the number of gears in the metabeam, and $e$ is the eccentricity. 

\renewcommand{\thefigure}{S8}
\begin{figure}[htb]
\centering
\includegraphics[scale=0.4]{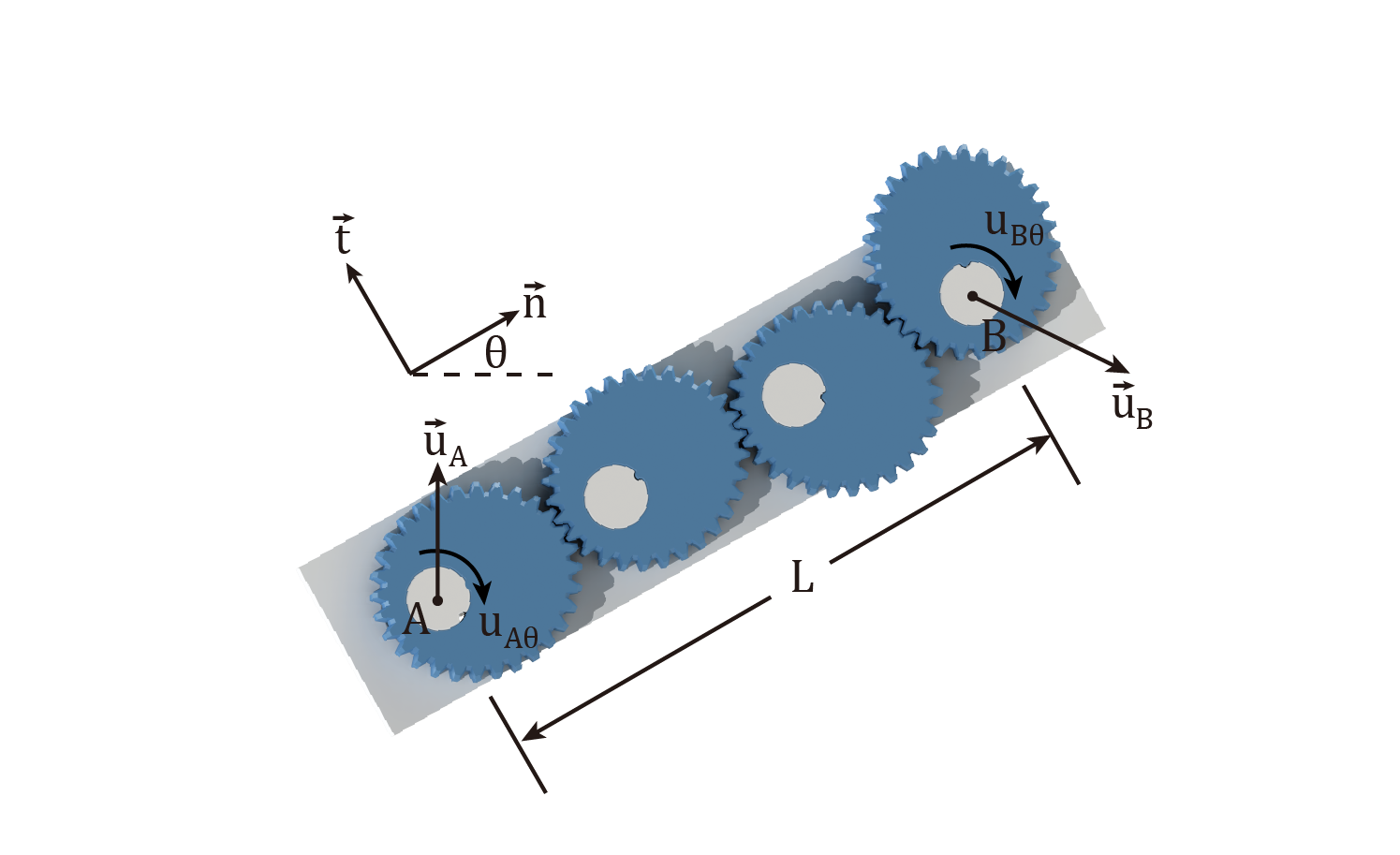}
\caption{Schematic illustration of the one-dimensional metabeam that assembles the two-dimensional generalized honeycomb lattice.
}\label{S7}
\end{figure}

An intuitive picture to visualize the term ``transmission rate", is to consider a metabeam that consists of two elliptic gears with contacting curvatures $r_1$ and $r_2$, respectively. This implies that the rotation angles of the gears are related by $\theta_2/\theta_1=-r_1/r_2=\lambda$, where $\lambda$ is referred to as the transmission rate along the chain, and the negative sign in the equation indicates that the adjacent gears rotate in opposite directions.

We now derive the analytic expression of the transmission rate in terms of $\theta$, namely the rotation angle of the bond orientation of the metabeam. To this end, we establish the highly nonlinear relationship between the rotation angles of the gears in the metabeam. We first consider a metabeam with only two gears. When the first gear rotates by an angle of $x$, the other gear rotates accordingly by an angle of $x+\Theta(x)$, where $\Theta(x)$ denotes the difference of the rotation angles between these two adjacent gears:
\begin{eqnarray}\label{guest_2}
\Theta(x) = {\rm sgn}\,(\sin x)\,\arccos\left(\frac{1-e^2+2e \cos x+2e^2 \cos^2 x}{1+e^2+2e\cos x}\right)-x.
\end{eqnarray}
It is intuitive the extend this result to the $N$-gear metabeam. The difference in the rotation angles between the last and first gear, is given by $\underbrace{\Theta\circ \Theta\circ \ldots \circ \Theta}_{N-1}(x)$, where $x$ is the angular displacement of the first gear. Next, we utilize Eq. (\ref{guest_2}) for Fig. \ref{S7}, where the rotation of the bond is also considered. We denote $\theta_1$ and $\theta_1^{(0)}$ as the initial and final bond orientations, respectively, and denote the angular displacements of the first and the $N$th gears, as $u_{1\theta}$ and $u_{N\theta}$, respectively. According to Eq. (\ref{guest_2}), these quantities yield the following highly nonlinear constraint,
\begin{eqnarray}\label{guest_1}
\theta_1-\theta_1^{(0)}-\underbrace{\Theta\circ \Theta\circ \ldots \circ \Theta}_{N-1}(\theta_1-\theta_1^{(0)}-u_{1\theta})=u_{N\theta }.
\end{eqnarray}
Eq. (\ref{guest_1}) allows us to express $u_{1\theta}$ in terms of $\theta_1$ and $u_{N\theta}$. Thus, the transmission rate $\lambda$ of the metabeam, which is defined as the differential ratio of the rotational speed between the first and last gears, is given by 
\begin{eqnarray}\label{guest_3}
\lambda(\theta_1) = \frac{d\overbrace{\Theta\circ \Theta\circ \ldots \circ \Theta}^{N-1}(x)}{dx}\bigg|_{x=u_{1\theta}(\theta_1-\theta_1^{(0)},u_{N\theta})-\theta_1+\theta_1^{(0)}}.
\end{eqnarray}
\renewcommand{\thefigure}{S9}
\begin{figure}[htb]
\centering
\includegraphics[scale=0.185]{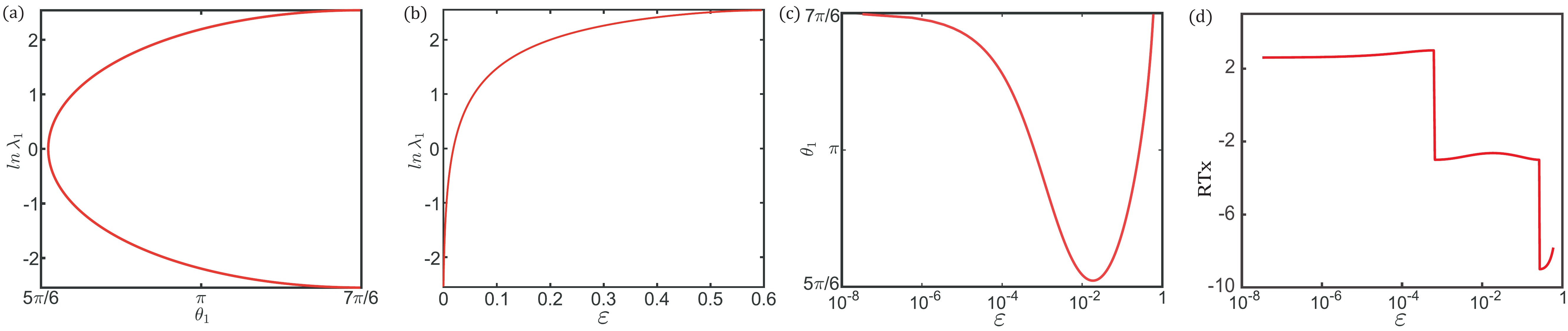}
\caption{In the metabeam with $N=4$ gears, we numerically compute the transmission rate $\lambda_1$, (a) as the function of the bond direction that varies from $\theta_1=5\pi/6$ to $7\pi/6$, and (b) as the function of the degree of nonlinearity in the metabeam, which varies from $\epsilon=0$ to $\epsilon=0.6$. (c) The relationship between the Guest mode angle and the degree of nonlinearity in the metabeams. (d) The relationship between the $x$-direction projection of the topological polarization $\bm{R}_{\rm T}$ and the degree of nonlinearity in the metabeams.
}\label{S21}
\end{figure}\\
We can further express the rotation angle of the first gear in the metabeam, $u_{1\theta}$, in terms of the degree of nonlinearity $\epsilon$ via the relationship $\epsilon = |s_e(u_{1\theta})/a(1-e)u_{1\theta}-1|$. Using Eq. (\ref{guest_1}), we express the angle of the bond, $\theta_1$, in terms of the degree of nonlinearity $\epsilon$. This bond angle will later serve as the Guest mode angle of the geared honeycomb lattice. Finally, the transmission rate of the metabeam, $\lambda$, can be expressed in terms of the degree of nonlinearity via Eq. (\ref{guest_3}). All these numerical results appear in Figs. \ref{S21}(b, c, d).

As we will demonstrate later, in the 2D honeycomb lattice that is subjected to the nonlinear shearing Guest mode, the angular displacements of the first and final gears, yield the relationship 
\begin{eqnarray}\label{guest_4}
u_{1\theta} = -u_{N\theta},\qquad\Rightarrow \qquad
\theta_1-\theta_1^{(0)}+u_{1\theta }-\underbrace{\Theta\circ \Theta\circ \ldots \circ \Theta}_{N-1}(\theta_1-\theta_1^{(0)}-u_{1\theta})=0.
\end{eqnarray}
Consequently, the angular displacement of the first gear, $u_{1\theta}$, is controlled by $\theta_1$ only. Finally, the transmission rate can be analytically derived in the following form: 
\begin{eqnarray}\label{guest_5}
\lambda(\theta_1) = \frac{d\overbrace{\Theta\circ \Theta\circ \ldots \circ \Theta}^{N-1}(x)}{dx}\bigg|_{x=u_{1\theta}(\theta_1-\theta_1^{(0)})-\theta_1+\theta_1^{(0)}}.
\end{eqnarray}
In Fig. \ref{S21}, we numerically compute the transmission rate $\lambda_1$ in the metabeam with $N=4$ gears, as the bond orientation $\theta_1$ varies. It is worth emphasizing that the non-injective mapping from $\theta_1$ to $\ln\lambda_1$ rightly demonstrates the highly nonlinear geometry and topological mechanics in every metabeam. As demonstrated by our Supplementary Video, when the uniform soft Guest mode finishes one cycle, the rotation angles in the gears, or equivalently speaking, the degree of nonlinearity in the metabeam, only finishes a half period. Our work differs from Ref.~\cite{vitelliPRX} in that when the Guest mode in the 2D honeycomb gear lattice reaches the point where the gears have rotated by $180^\circ$ relative to the beams they are mounted on, the lattice geometry is completely restored to its initial configuration. When the Guest mode angle completes two periods, the gears in the metabeams complete a full $360^\circ$ rotation and return to their initial configurations.  

\renewcommand{\thefigure}{S10}
\begin{figure}[htb]
\centering
\includegraphics[scale=0.8]{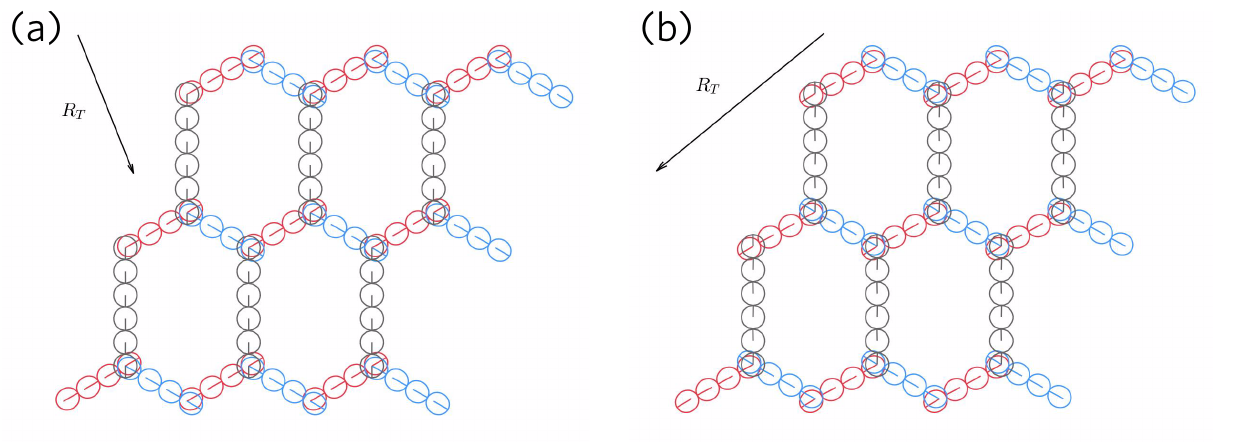}
\caption{In (a) and (b), the bond orientations of the lattice structures are the same, but the gear orientations in (b) are rotated by 180 degrees compared to those in (a).
}\label{S30}
\end{figure}

As shown by the red, blue, and green metabeams in Fig. \ref{S30}, the gear orientations in (a) and (b) are mirror images of each other. When these metabeams are assembled in the 2D structure, we are allowed to construct two honeycomb lattices, as shown in Fig. \ref{S30}(a) and (b), whose bond orientations are identical, but their gear orientations are mirror images of one another. Thus, starting from Fig. \ref{S30}(a), when the Guest mode finishes one cycle to Fig. \ref{S30}(b) and makes the lattice restore to its initial configuration, all gear orientations are guaranteed to reach the state of 180-degree rotation. This result offers the intuitive picture of why one lattice configuration corresponds to two distinct topological mechanics in the metabeams.

Now we allow the metabeam to be freely rotatable in 2D plane, where the rotation angle is $\delta\theta = (\bm{u}_b-\bm{u}_a)\cdot\hat{t}/L$, and the small-displacement approximation has been used here. Together with the metabeam, all gears uniformly rotate by $\theta$, from which we build the transverse constraint $e_t = \lambda(u_{a\theta}-\delta\theta)+(u_{b\theta}-\delta\theta) = (1+\lambda)(\bm{u}_a-\bm{u}_b)\cdot \hat{t}/L+\lambda u_{a\theta}+u_{b\theta}$. We summarize these results to derive the ``compatibility matrix" of a single metabeam,
\begin{eqnarray}
(e_n,e_t)^\top
=
(g_1(\theta,\lambda, L),g_2(\theta,\lambda,L))
(u_{ax},u_{ay},u_{a\theta},u_{bx},u_{by},u_{b\theta})^\top
\end{eqnarray}
where $\top$ denotes matrix transpose, and
\begin{eqnarray}\label{g}
g_1( \theta ,\lambda ,L) = \left( {\begin{array}{*{20}{c}}
-\cos\theta & -\sin\theta & 0\\
\frac{1+\lambda}{L}\sin\theta & -\frac{1+\lambda}{L}\cos\theta & \lambda 
\end{array}} \right),\qquad
g_2(\theta ,\lambda ,L) = \left( \begin{array}{*{20}{c}}
\cos \theta & \sin \theta & 0\\
-\frac{1+\lambda}{L}\sin \theta & \frac{1 +\lambda}{L}\cos\theta & 1
\end{array} \right)\qquad
\end{eqnarray}
are $2\times 3$ matrices that are later used to construct the compatibility matrix of the generalized honeycomb lattice.

We continue our discussion by considering the generalized honeycomb lattice, in which every unit cell is composed of two sites $A$ and $B$, and three metabeams labelled by 1, 2, and 3, as shown in Fig. \ref{S6}. The beam orientations are denoted as $\hat{n}_i = (\cos\theta_i, \sin\theta_i)$, the transmission rates are $\lambda_i$, and the beam lengths are $L_i$ for $i=1,2,3$. Nearest-neighbor beams are joined by gears in the vertical direction, where a vertical hinge prevents the top and bottom gears from relative rotations and displacements. 
The primitive vectors are denoted as $\bm{a}_1$ and $\bm{a}_2$, and the reciprocal vectors are defined as $\bm{b}_1 = 2\pi \bm{a}_2\times \bm{a}_3/\bm{a}_1\cdot(\bm{a}_2\times \bm{a}_3)$ and $\bm{b}_2 = 2\pi \bm{a}_3\times \bm{a}_1/\bm{a}_1\cdot(\bm{a}_2\times \bm{a}_3)$, where $\bm{a}_3 = (0,0,1)$. We label the unit cell at the position $\bm{R} = n_1 \bm{a}_1+n_2\bm{a}_2$ as $\bm{n} = (n_1,n_2)$. Following this convention, the sites are labelled as $A(\bm{n})$ and $B(\bm{n})$. 
The site displacements are in consequence denoted as $(u_{Ax}(\bm{n}), u_{Ay}(\bm{n}), u_{A\theta}(\bm{n}))$ and $(u_{Bx}(\bm{n}), u_{By}(\bm{n}), u_{B\theta}(\bm{n}))$, and the longitudinal and transverse elongations of the beam are denoted as $e_{ni}(\bm{n})$ and $e_{ti}(\bm{n})$ for $i=1,2,3$.


As topological floppy modes arise from isostatic/Maxwell lattices, we have to count the numbers of constraints versus degrees of freedom in order to confirm isostaticity. Every node has three degrees of freedom, namely two translational modes, and one rotational mode induced by the gear rotation. Each metabeam imposes two constraints (i.e., the longitudinal and transverse constraints) on the two connected sites, thereby offering $2/2=1$ constraint per site. Since the coordination number of honeycomb lattice is $z=3$, the lattice in consequence has the balanced degrees of freedom and constraints to stay at the isostatic/Maxwell point.

As the underlying system is constructed from spatially repetitive frames, the aforementioned isostatic condition can be alternatively confirmed in the perspective of unit cells. Each unit cell has two sites to host $2\times 3 = 6$ degrees of freedom, and three metabeams to provide $3\times 2 = 6$ constraints, to stay at the isostatic point. 

Cutting the lattice boundaries into open grants excess degrees of freedom which eventually manifest as exponentially localized zero-frequency ``floppy" modes. Their topological attributes lie in the analysis of compatibility matrix $\bm{C}$, which is a linear matrix operator that maps the site movements to the beam extensions via $\bm{C}\bm{u} = \bm{e}$. Here, $\bm{u} = (\ldots, u^\top(\bm{n}),\ldots)^\top$ denotes the displacement field of all sites with $\bm{u}(\bm{n}) = (u_{Ax}(\bm{n}), u_{Ay}(\bm{n}), u_{A\theta}(\bm{n}), u_{Bx}(\bm{n}), u_{By}(\bm{n}), u_{B\theta}(\bm{n}))^\top$ the unit cell displacement, $\bm{e} = (\ldots, e(\bm{n}),\ldots)$ stands for all beam extensions with $e(\bm{n}) = (e_{n1}(\bm{n}), e_{t1}(\bm{n}), e_{n2}(\bm{n}), e_{t2}(\bm{n}), e_{n3}(\bm{n}), e_{t3}(\bm{n}))$ the beam elongations of the unit cell. 

Spatially periodic frames enjoy the nice property that any mechanical modes can be Fourier-transformed into momentum space through $e(\bm{n}) = \sum_{\bm{k}}e(\bm{k})e^{\mathrm{i}\bm{k}\cdot \bm{R}(\bm{n})}$ and $u(\bm{n}) = \sum_{\bm{k}}u(\bm{k})e^{\mathrm{i}\bm{k}\cdot \bm{R}(\bm{n})}$, where $\bm{k}$ is the two-dimensional wavevector. 
As such, the lattice compatibility matrix is reduced to a $6\times 6$ block-diagonal matrix in momentum space that yields 
$\bm{C}(\bm{k})\bm{u}(\bm{k}) = \bm{e}(\bm{k})$, where 
\begin{eqnarray}\label{compatibility2}
\bm{C}(\bm{k}) = \left( \begin{array}{*{20}{c}}
e^{\mathrm{i}\bm{k}\cdot\bm{a}_1}g_1(\theta_1,\lambda_1,L_1) & g_2( \theta_1,\lambda_1,L_1 )\\
g_1(\theta_2,\lambda_2,L_2) &  g_2(\theta_2,\lambda_2,L_2)\\
g_1(\theta_3,\lambda_3,L_3) & e^{\mathrm{i}\bm{k} \cdot (\bm{a}_2-\bm{a}_1)}{g_2}(\theta_3,\lambda_3,L_3)
\end{array} \right).
\end{eqnarray}
To be explicit, we incorporate all functions in Eq. (\ref{g}) into the compatibility matrix of Eq. (\ref{compatibility2}). Due to the nature of the nonlinear soft shearing Guest mode, the rotation angles of the gears on the type-3 metabeam yield the relationship $u_{A\theta}=-u_{B\theta}$. As a result, the gear rotations on type-1 and type-2 metabeams in the honeycomb lattice, satisfy the constraint $u_{A\theta}=-u_{B\theta}$ as well. This constraint allows us to simplify the transmission rate from Eq. (\ref{guest_3}) to (\ref{guest_5}). In addition, we use the relationships $\theta_2=\theta_1-\pi$, $\theta_3=\pi/2$, $\lambda_2=\lambda_1^{-1}$, and $\lambda_3=1$, to express the compatibility matrix in terms of the Guest mode angle $\theta_1$, which is the bond orientation of the first-type metabeam: 
\begin{eqnarray}\label{compatibility}
 & {} & \bm{C}(\bm{k})=\nonumber \\
 & {} &  \left( \begin{array}{*{20}{c}}
-e^{\mathrm{i}\bm{k}\cdot\bm{a}_1}\cos\theta_1 & -e^{\mathrm{i}\bm{k}\cdot\bm{a}_1}\sin\theta_1 & 0 & \cos\theta_1 & \sin\theta_1 & 0 \\
e^{\mathrm{i}\bm{k}\cdot\bm{a}_1}\frac{1+\lambda_1(\theta_1)}{L_1}\sin\theta_1 & -e^{\mathrm{i}\bm{k}\cdot\bm{a}_1}\frac{1+\lambda_1(\theta_1)}{L_1}\cos\theta_1 & e^{\mathrm{i}\bm{k}\cdot\bm{a}_1}\lambda_1(\theta_1) & -\frac{1+\lambda_1(\theta_1)}{L_1}\sin\theta_1 & \frac{1+\lambda_1(\theta_1)}{L_1}\cos\theta_1 & 1\\
\cos\theta_1 & \sin\theta_1 & 0 & -\cos\theta_1 & -\sin\theta_1 & 0 \\
-\frac{1+\lambda_1(\theta_1)}{\lambda_1(\theta_1) L_1}\sin\theta_1 & \frac{1+\lambda_1(\theta_1)}{\lambda_1(\theta_1) L_1}\cos\theta_1 & \frac{1}{\lambda_1(\theta_1)} & \frac{1+\lambda_1(\theta_1)}{\lambda_1(\theta_1) L_1}\sin\theta_1 & -\frac{1+\lambda_1(\theta_1)}{\lambda_1(\theta_1) L_1}\cos\theta_1 & 1\\
0 & -1 & 0 & 0 & e^{\mathrm{i}\bm{k}\cdot(\bm{a}_2-\bm{a}_1)} & 0 \\
\frac{2}{L_3} & 0 & 1 & -e^{\mathrm{i}\bm{k}\cdot(\bm{a}_2-\bm{a}_1)}\frac{2}{L_3} & 0 & e^{\mathrm{i}\bm{k}\cdot(\bm{a}_2-\bm{a}_1)}\\
\end{array} \right).\nonumber \\
\end{eqnarray}
where the transmission rate in the metabeam, namely $\lambda_1(\theta_1)$, is nonlinearly controlled by the Guest mode angle $\theta_1$. Here, the nonlinear functional dependence of $\lambda_1$ on $\theta_1$ is pictorially depicted in the following Fig. \ref{S21}(a), with its analytical expression captured by Eq. (\ref{guest_5}). 
\renewcommand{\thefigure}{S11}
\begin{figure}[htb]
\centering
\includegraphics[scale=0.25]{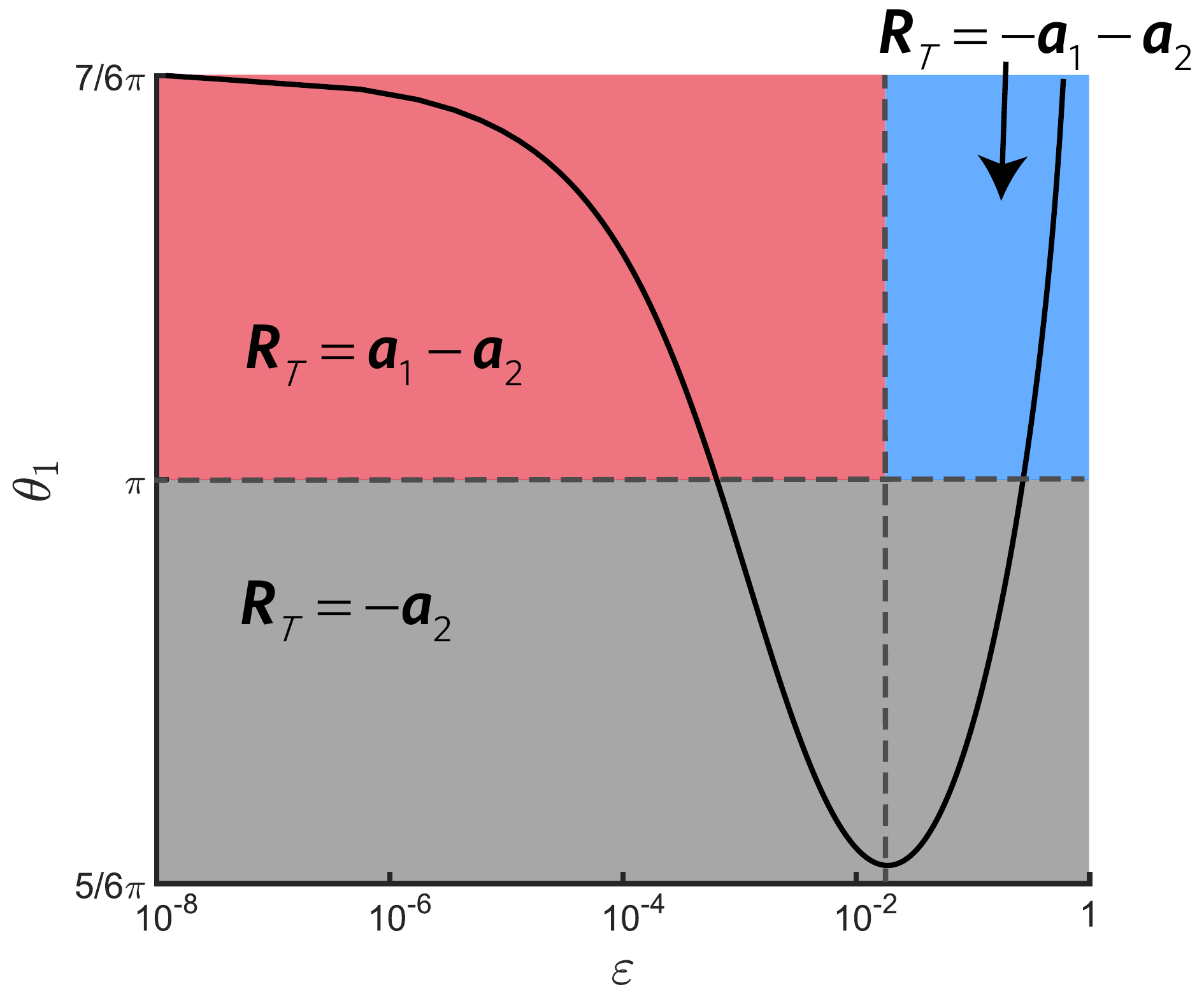}
\caption{Phase diagram of the topological polarization in terms of the Guest mode angle $\theta_1$ and the degree of nonlinearity, $\epsilon$, in the metabeams.
}\label{S20}
\end{figure}

The topological characterization of the lattice is mathematically formulated by the topological polarization,
\begin{eqnarray}\label{polarization}
\bm{R}_{\rm T} = \sum_{i = 1,2} \mathcal{N}_i \bm{a}_i,
\end{eqnarray}
where $\mathcal{N}_i$ is the winding number defined in Brillouin zone
\begin{eqnarray}\label{wind}
\mathcal{N}_i=-\frac{1}{2\pi}\oint_{C_i} d\bm{k} \cdot \nabla_{\bm{k}}\,\mathop{\rm Im}\nolimits \ln \det \bm{C}(\bm{k}),
\end{eqnarray}  
and $C_i=\bm{k}\to \bm{k}+\bm{b}_i$ is a closed path in Brillouin zone parallel to the reciprocal vector $\bm{b}_i$. Fig. \ref{S20} plots the topological phase diagram of the polarization in terms of the Guest mode angle $\theta_1$ and the corresponding degree of nonlinearity $\epsilon$ in the type-1 metabeam.

We cut the lattice from PBCs to mixed boundary conditions, where OBC takes place in $\bm{a}_1$ and PBC is retained in $\bm{a}_2$. The missing beams on open boundaries grant excess degrees of freedom, which are manifested as topologically polarized floppy modes on lattice boundaries. It is notable that topological polarization $\bm{R}_{\rm T}$ alone is not sufficient to compute the number of localized floppy modes. Instead, the ``local polarization" $\bm{R}_{\rm L}$ and the topological polarization $\bm{R}_{\rm T}$ together [4,5] determine the number of topological floppy modes through 
\begin{equation}
\nu=\frac{1}{2\pi}\vec G\cdot(\bm{R}_{\rm T}+\bm{R}_{\rm L}), 
\end{equation}
where $\nu$ denotes the number of localized floppy modes on the open boundary, and $\vec G$ is the reciprocal lattice vector pointing outward to the normal of the cutting edge. The calculation of $\bm{R}_{\rm L}$ is elaborated as follows. First, we take the unit cell colored in red in Fig. \ref{S6} as the reference unit cell, with $\bm{R}_A^{(0)}$, $\bm{R}_B^{(0)}$ the site positions and $\bm{R}_i^{(0)}$ for $i=1,2,3$ the positions of the metabeams. Other choices of unit cell can be quantitatively characterized by a dipole moment that depicts the different positions relative to the reference unit cell, via a dipole moment vector 
\begin{equation}
\bm{R}_{\rm L} = 3(\bm{R}_A-\bm{R}_A^{(0)})+3(\bm{R}_B-\bm{R}_B^{(0)})-2\sum_{i=1,2,3}(\bm{R}_i-\bm{R}_i^{(0)}),
\end{equation}
where $\bm{R}_A$, $\bm{R}_B$ are the site positions and $\bm{R}_i$ for $i=1,2,3$ are the beam positions of the other choices of the unit cell. The factor of 3 indicates that each site contains three degrees of freedom, and the factor of 2 indicates the two constraints provided by each beam.

The local polarization $\bm{R}_{\rm L}$ has to be introduced due to the following reason. As the compatibility matrix can change upon the gauge choice of the unit cell, the topological polarization $\bm{R}_{\rm T}$ is gauge-dependent, which is in stark contrast to the gauge-invariant distribution of topological floppy modes. Thus, the local polarization $\bm{R}_{\rm L}$ is introduced to cancel the gauge dependence of $\bm{R}_{\rm T}$. $\bm{R}_{\rm T}$ and $\bm{R}_{\rm L}$ together address the gauge-invariant distribution of topological boundary floppy modes. 

Given the topological polarization $\bm{R}_{\rm T} = \mathcal{N}_1\bm{a}_1+\mathcal{N}_2\bm{a}_2$, we compute $\bm{R}_{\rm L}$ to determine the number of floppy modes on the open boundaries of Fig. \ref{S5}. On the left open boundary, the outward normal vector is $\vec G_{\rm left}=-\bm{b}_1$. The unit cell compatible to the left open boundary is colored by red, which is identical to our reference unit cell. Thus, we have $\bm{R}_1^{(\rm left)}=\bm{R}_1^{(0)}-\bm{a}_1$, $\bm{R}_2^{(\rm left)}=\bm{R}_2^{(0)}$, $\bm{R}_3^{(\rm left)}=\bm{R}_3^{(0)}$, and $\bm{R}_A^{(\rm left)}=\bm{R}_A^{(0)}$, $\bm{R}_B^{\rm left}=\bm{R}_B^{(0)}-\bm{a}_1$, to enable $\bm{R}_{\rm L}=-\bm{a}_1$ for the left open boundary. The number of topological floppy modes is $\nu_{\rm left}=-\mathcal{N}_1 + 1$. On the right boundary, the outward normal vector reads $\vec G_{\rm right}=+\bm{b}_1$. Compatible to the right open boundary, the unit cell is depicted by the blue one. This new unit cell offers $\bm{R}_1^{(\rm right)}=\bm{R}_1^{(0)}$, $\bm{R}_2^{(\rm right)}=\bm{R}_2^{(0)}$, $\bm{R}_3^{(\rm right)}=\bm{R}_3^{(0)}+\bm{a}_1$, and $\bm{R}_A^{(\rm right)}=\bm{R}_A^{(0)}+\bm{a}_1$, $\bm{R}_B^{\rm right}=\bm{R}_B^{(0)}$, to enable $\bm{R}_{\rm L} = +\bm{a}_1$. Thus, the floppy mode number of the right open boundary is $\nu_{\rm right}=\mathcal{N}_1+1$.

As we continuously change the geometry of the generalized honeycomb lattice via the Guest mode~\cite{guest2003determinacy}, the system experiences discontinuous jumps in the topological invariant with $\mathcal{N}_1=1,0,-1$. In these topologically distinct phases, the count of topological boundary floppy modes are $(\nu_{\rm left},\nu_{\rm right})=(0,2)$ (left boundary rigid and right boundary soft), $(\nu_{\rm left},\nu_{\rm right})=(1,1)$ (both boundaries are soft), and $(\nu_{\rm left},\nu_{\rm right})=(2,0)$ (left boundary soft and right boundary rigid), as shown in Figs. (2) and (3) of the main text.

The real-space distribution of floppy modes is calculated for each metabeam $\alpha$ defined as
\begin{eqnarray}
\eta_\alpha = \frac{1}{2N_0}\sum_{s=1}^{N_0}\left(|\bm{u}_{\alpha,1}^{(s)}|^2+|\bm{u}_{\alpha,2}^{(s)}|^2\right)
\end{eqnarray}
where the sum is over all floppy modes labeled by $s$, and $\bm{u}_{\alpha,i=1,2}^{(s)}$ denotes the $s$-th floppy mode displacements on the two ends of the considered metabeam $\alpha$. This distribution is normalized by the total number of floppy modes $N_0$, as pictorially reflected in Figs. 2(d-g) and Fig. 4(a,c) of the main text. 

\renewcommand{\thefigure}{S12}
\begin{figure}[htb]
\centering
\includegraphics[scale=0.5]{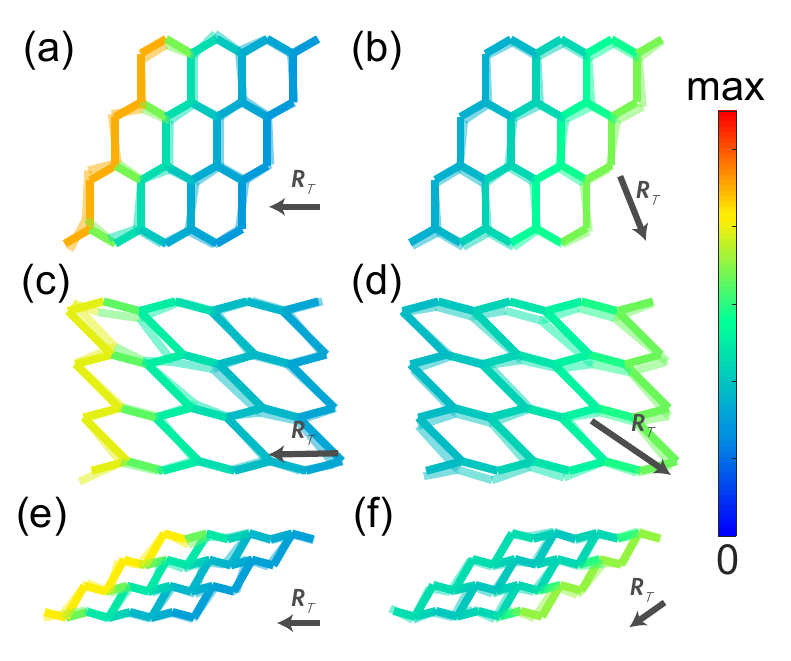}
\caption{Different choices of gear parameters and lattice geometries that produce polarized topological boundary floppy modes. The transmission rates and bond orientations in each metabeam, namely $(\lambda_1,\lambda_2,\lambda_3,\theta_1,\theta_2,\theta_3)$, are given by: $\left( 0.16,0.08,1,\frac{7\pi}{6}, \frac{11\pi}{6},\frac{\pi}{2} \right)$ in (a), $\left( 0.16,0.21,0.16,\frac{7\pi}{6},\frac{11\pi}{6},\frac{\pi}{2} \right)$ in (b), $\left( 0.16,0.08,1,\frac{13\pi}{12},\frac{23\pi}{12},\frac{3\pi}{4} \right)$ in (c), $\left( 0.1,1.08,1,\frac{13\pi}{12},\frac{23\pi}{12},\frac{3\pi}{4} \right)$ in (d), $\left( 0.08,0.21,1,\frac{11\pi}{12},\frac{\pi}{12},\frac{\pi}{3} \right)$ in (e), and $\left( 3.4,0.09,0.21,\frac{11\pi}{12},\frac{\pi}{12},\frac{\pi}{3} \right)$ in (f).
}\label{S13}
\end{figure}

In Fig.2(a) of the main text, it is convenient to use 3D-printed elliptical gears that are identical for both type-1 and type-2 metabeams. However, other choices of elliptical gears are also feasible for achieving the same topological phase of the 2D lattice, as long as type-1 and type-2 metabeams are polarized in the upper-right and lower-right directions, respectively. For simplicity, circular gears are used for type-3 metabeams, resulting in an overall rightward polarization of the lattice. Alternatively, choosing elliptical gears with $ \lambda \ne 1$ for type-3 metabeams, as demonstrated in Fig. \ref{S13}(b), still ensures the topological polarization that points toward the lower-right corner.

By varying all eccentricities of the elliptic gears, we conduct a comprehensive study of all topological phase diagrams in Figs. \ref{phaseDiagram1}, Fig. \ref{phaseDiagram2}, and Fig. \ref{phaseDiagram3}, which reveal a total of 7 different topological phases and mechanical properties in the two-dimensional mechanical metamaterial. These phase diagrams are obtained by varying all eccentricities, namely $e_1$, $e_2$, and $e_3$, and solving the lattice configurations along the nonlinear Guest mode. The eccentricity $e_1$ is $e_1=0.3$, $e_1=0.4$, and $e_1=0.5$ for Figs. \ref{phaseDiagram1}, \ref{phaseDiagram2}, and \ref{phaseDiagram3}, respectively, while the other two parameters, namely $e_2$ and $e_3$, vary from $-0.5$ to $0.5$ in these figures.

The experimental prototype we used has $(e_1,e_2,e_3)=(0.4,0.4,0)$, whose phase diagram (Fig. 2(c) of the main text) is shown in the fourth row, second column of the table in Fig. \ref{phaseDiagram2}. While such specific choices of eccentricities used in our experiments were motivated by convenience in 3D-printing, we found that a range of values, such as $(e_1,e_2,e_3) =$ (0.3, 0.5, -0.3), (0.3, 0.4, -0.3), (0.3, 0.5, 0), (0.3, 0.4, 0), (0.3, 0.3, 0) in Fig. \ref{phaseDiagram1}, $(e_1,e_2,e_3) =$ (0.4, 0.5, -0.4), (0.4, 0.5, -0.3), (0.4, 0.4, -0.3), (0.4, 0.5, 0), (0.4, 0.3, 0) in Fig. \ref{phaseDiagram2}, and $(e_1,e_2,e_3) =$ (0.5, 0.5, -0.3), (0.5, 0.4, -0.3), (0.5, 0.5, 0), (0.5, 0.4, 0), (0.5, 0.3, 0) in Fig. \ref{phaseDiagram3}, can also result in similar topological phases and nonlinear mechanical responses (see colors in the corresponding phase diagrams). These findings exemplify the topological robustness of the mechanical properties in our metamaterial. 

When the Guest mode shears the honeycomb lattice, it changes the orientation of each metabeam, causing the gears in the metabeams to rotate and inducing a nonlinear topological phase transition in their mechanical properties. The positions of softness and rigidity switch in each metabeam, leading to a state where all of the honeycomb's floppy modes move from one open boundary to the opposite. As shown in Eq. (\ref{compatibility}), the highly nonlinear mechanical properties of the metabeams are incorporated into the compatibility matrix, where the highly nonlinear ``transmission rates" $\lambda_1$ and $\lambda_2$ vary with the Guest mode angle, leading to a complete reversal of the topological polarization in the honeycomb lattice as the Guest mode angle changes, and the reversal of the stiffness contrast as well.

This remarkable effect sets the honeycomb lattice apart from previous works, such as~\cite{vitelliPRX, xiu2023synthetically, PhysRevB.101.104106}. Unlike these works, which achieve only a partial exchange of mechanical floppy modes and states of self-stress, the geared honeycomb lattice exploits the embedded nonlinear topological mechanical properties of each metabeam to achieve a complete reversal of stiffness contrast. This unique mechanical property makes the material a candidate for flexible topological metamaterials with high tunability.

\renewcommand{\thefigure}{S13}
\begin{figure}[htb]
\centering
\includegraphics[scale=0.55]{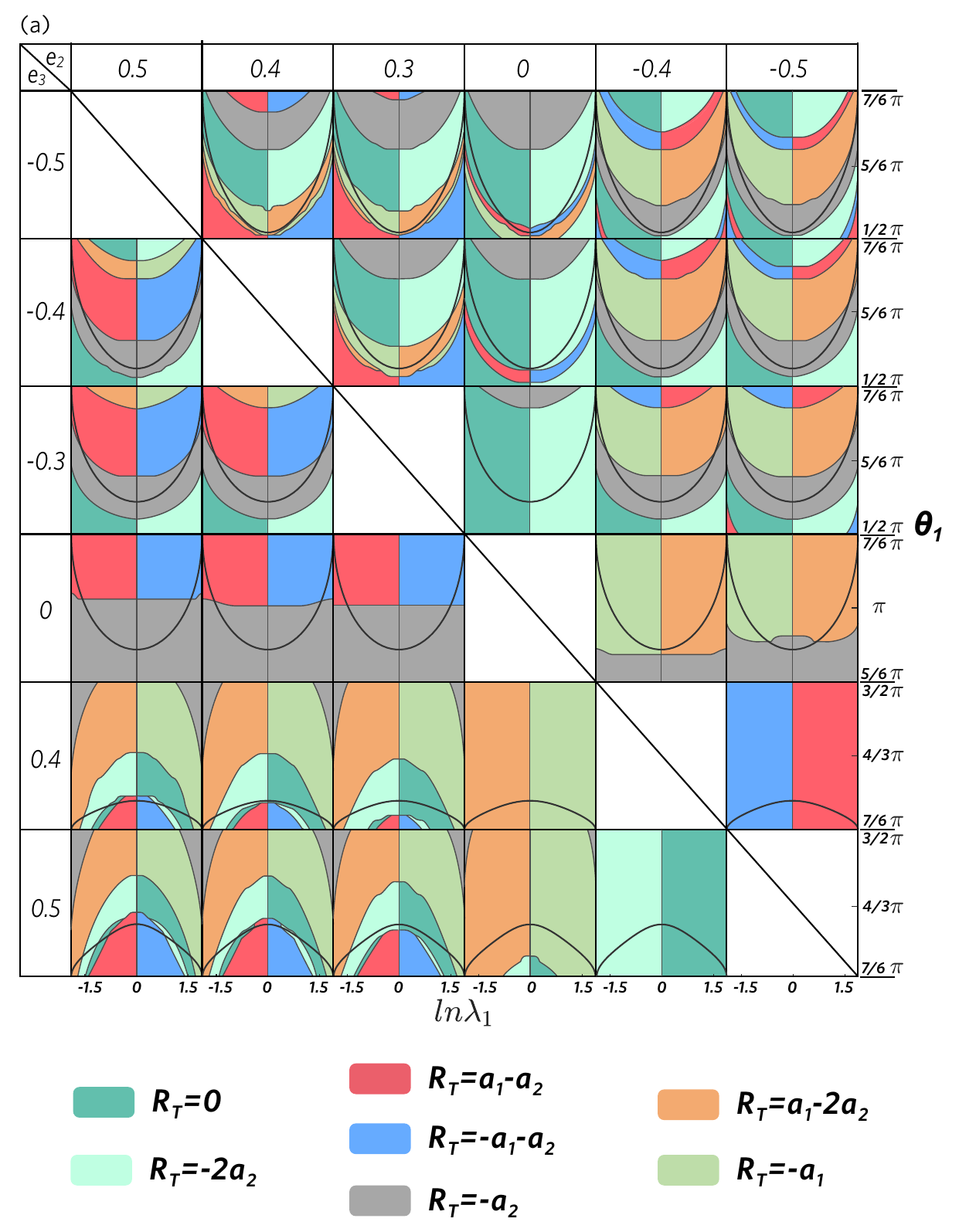}
\caption{A comprehensive table of the topological phase diagrams of our two-dimensional geared metamaterial. The phase diagrams were obtained by letting $e_1=0.3$, and varying the eccentricities $e_2$  and $e_3$ of the elliptic gears in the metamaterial. The lattice topology on the diagonal elements of the table is undefined due to a gapless phonon band structure that occurs when $e_2=e_3$. 
}\label{phaseDiagram1}
\end{figure}

\renewcommand{\thefigure}{S14}
\begin{figure}[htb]
\centering
\includegraphics[scale=0.55]{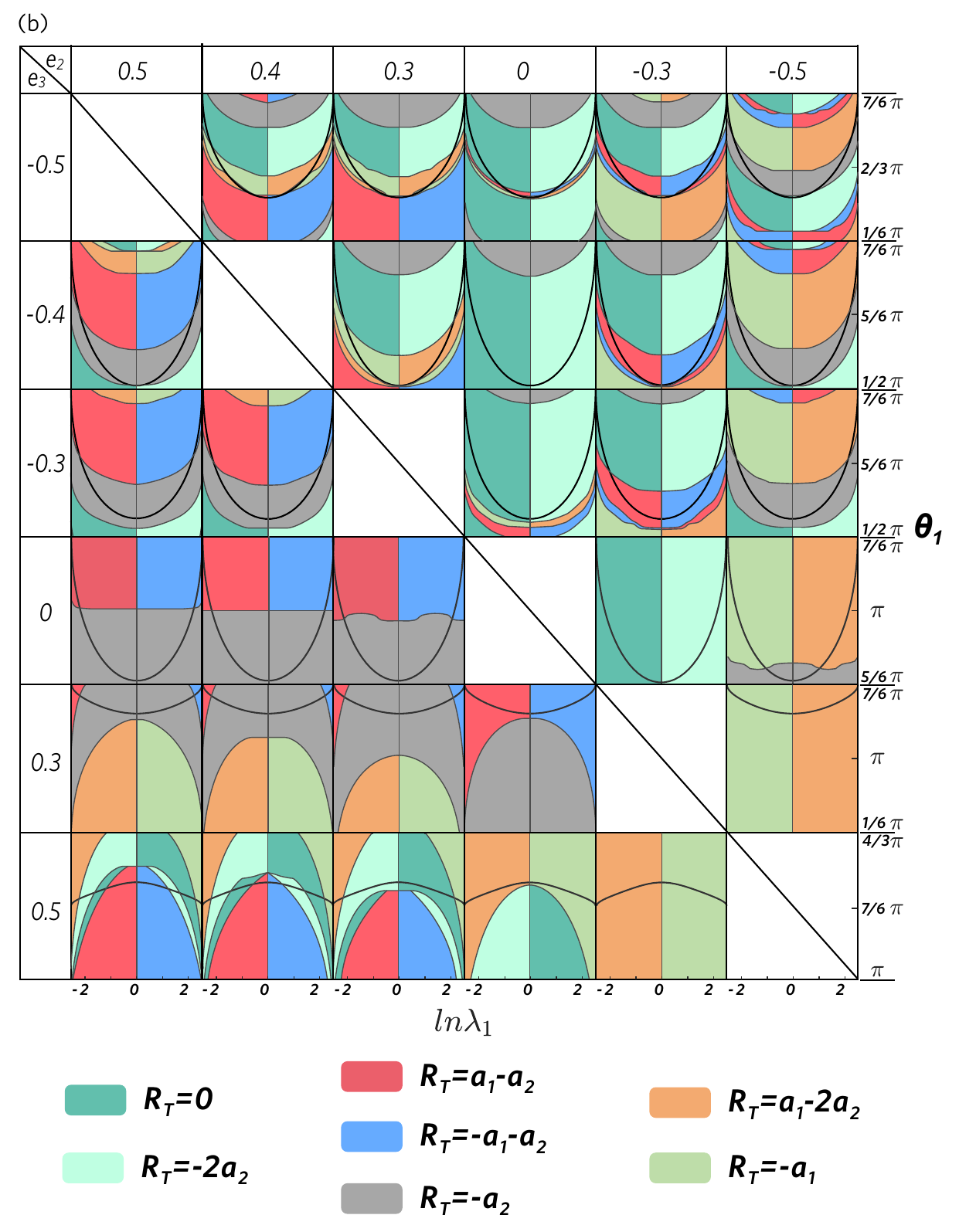}
\caption{The table of topological phase diagrams of the two-dimensional metamaterial, which is obtained by letting $e_1=0.4$, and varying the eccentricities $e_2$  and $e_3$ of the elliptic gears.
}\label{phaseDiagram2}
\end{figure}

\newpage
\renewcommand{\thefigure}{S15}
\begin{figure}[htb]
\centering
\includegraphics[scale=0.55]{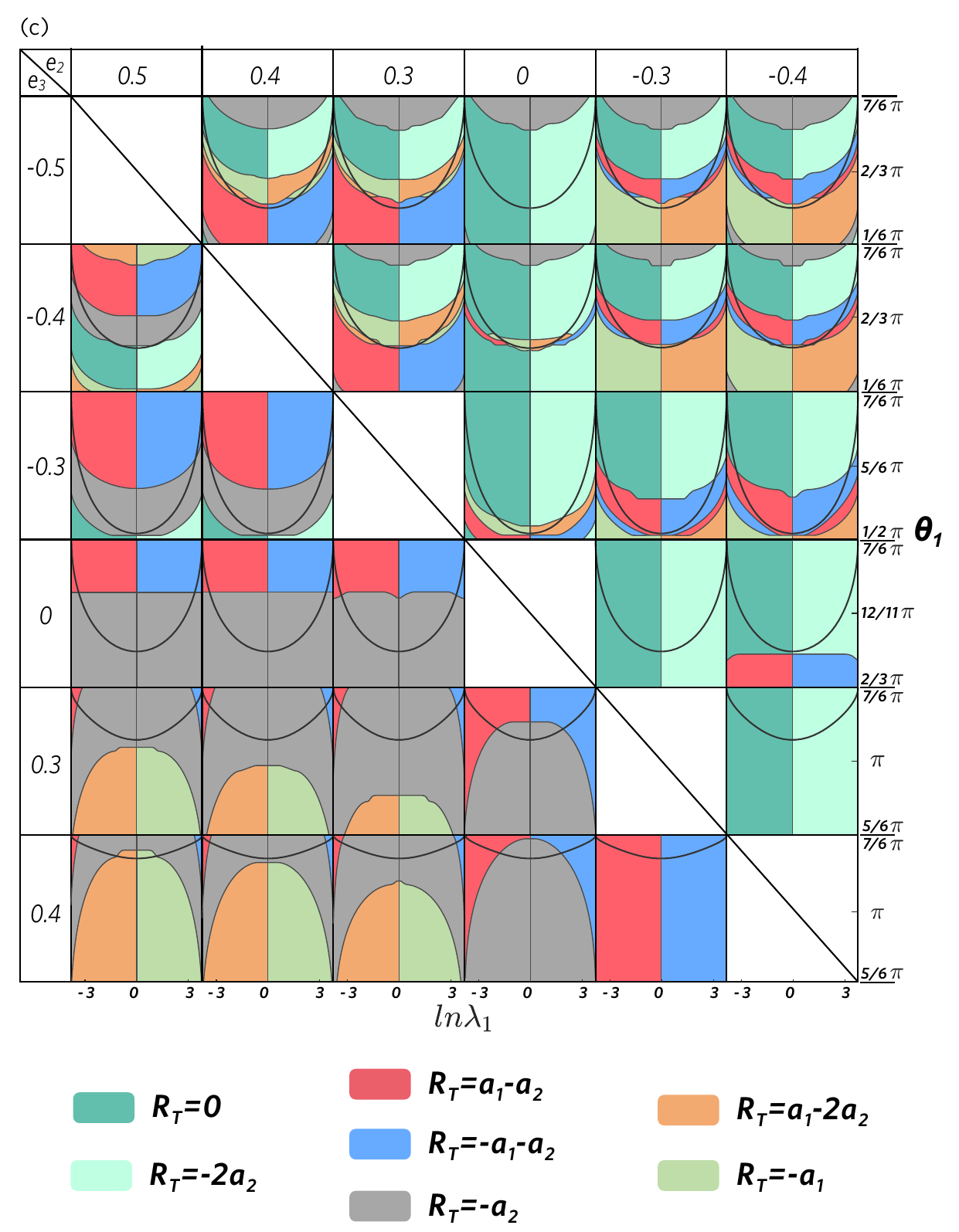}
\caption{The table of topological phase diagrams of the 2D metamaterial, which is obtained by letting $e_1=0.5$, and varying the eccentricities $e_2$  and $e_3$.
}\label{phaseDiagram3}
\end{figure}

\subsection{Boundary stiffness measurements of the geared honeycomb lattice}

\renewcommand{\thefigure}{S16}
\begin{figure}[htb]
\centering
\includegraphics[scale=0.75]{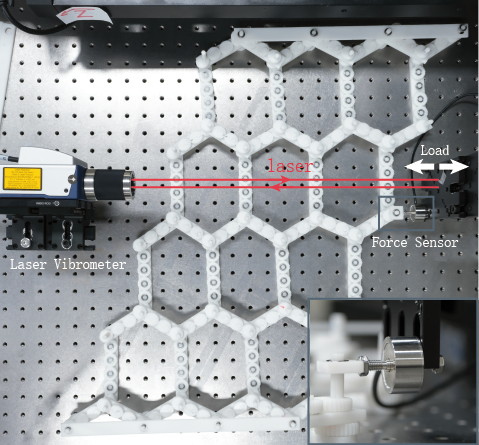}
\caption{The experimental setup for the measurement of the boundary stiffness in the elliptically geared honeycomb lattice.
}\label{S8}
\end{figure}

Fig. \ref{S8} shows the experimental setup to measure the boundary stiffness of the geared honeycomb lattice. We place the 2D geared metamaterial on a smooth platform to perform the force-displacement measurements of the left and right boundaries. The measured boundary is left open, whereas the other three boundaries are fixed. The boundary force is measured by a mechanical sensor, with one end mounted on a platform that moves horizontally at a constant speed, and the other end attached to the open boundary. A laser vibrometer vertically illuminates the mobile platform to read the displacement of the sensor. We record the force-displacement curves as the mobile platform pushes and pulls the boundary. 

\subsection{Internal structure of co-axial gears}

\renewcommand{\thefigure}{S17}
\begin{figure}[htb]
\centering
\includegraphics[scale=0.75]{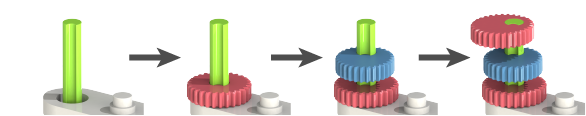}
\caption{The manufacturing details of the co-axial gears.}
\label{S9}
\end{figure}

To eliminate the relative displacement among co-axial gears, we first carve grooves on the pivot, and 3D print the corresponding humps on the hollow focal points to match these grooves. The pivot penetrates through the hollow focal points of all three elliptic gears, which prevents the relative horizontal displacements among these hollow focal points. As a result, relative rotations between the gears and the pivots are forbidden by the conjugated concave grooves and convex humps, which further prevents the co-axial gears from relative rotations. We have enlarged the grooves and humps in Fig. \ref{S9} and the video to make them more visible.

\subsection{The manual process of beam rearrangement and lattice reconfiguration}

\renewcommand{\thefigure}{S18}
\begin{figure}[htb]
\centering
\includegraphics[scale=0.6]{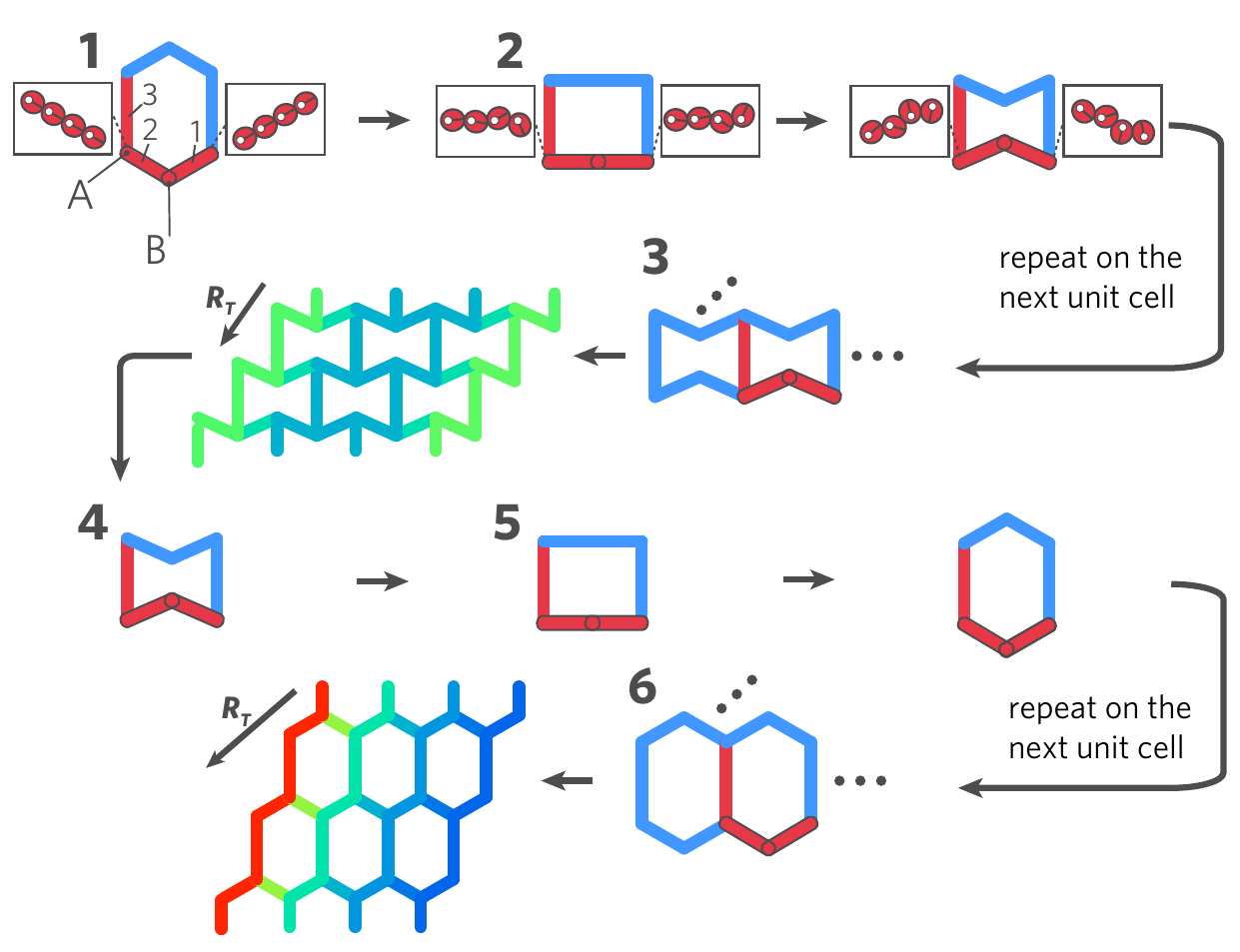}
\caption{The manual process of beam rearrangement and lattice reconfiguration.}
\label{S10}
\end{figure}

Our lattice is at the isostatic point~\cite{guest2003determinacy},which allows us to reconfigure the structure in two ways : (1) by manually rearranging the beams and allowing the gear rotations to follow, or (2) by driving the gear rotations and having the beam arrangements follow. Both methods result in the same reconfiguration.

Fig. \ref{S10} elaborates the manual reconfiguration that includes a topological phase transition of the 2D lattice. In Fig. \ref{S10}(1), the topological polarization $\bm{R}_{\rm T}$ points towards the lower-right corner of the lattice and indicates the topological floppy modes on the right open boundary. In Fig. \ref{S10}(2), we select a B-site in a unit cell of the honeycomb lattice. We slowly rotate the coaxial gears at this site-B in a clockwise way. The supporting beams rotate following these gears, with beam 1 rotating clockwise, beam 2 rotating counterclockwise, and beam 3 remaining non-rotational. Subsequently, Fig. \ref{S10}(3) illustrates the repetition of this procedure for all unit cells until the entire structure reaches an auxetic lattice configuration. As shown in Fig. \ref{S10}(4), we proceed to rotate the coaxial gears gradually at site-B of the unit cell in a clockwise direction. Meanwhile, beam 1 experiences counterclockwise rotation, beam 2 undergoes clockwise rotation, and beam 3 remains non-rotational. We repeat this process for all unit cells, as shown in Fig. \ref{S10}(6), until the overall deformation is accomplished. In Fig. \ref{S10}(6), the beam orientations remain the same as those in Fig. \ref{S10}(1), but the gear orientations are completely reversed. Consequently, the topological polarization $\bm{R}_{\rm T}$ undergoes a transition and points towards the lower left corner, indicating that the topological floppy modes are now localized on the left open boundary of the lattice.

\subsection{Guest modes in the presence of dislocations}

\renewcommand{\thefigure}{S19}
\begin{figure}[htb]
\centering
\includegraphics[scale=0.73]{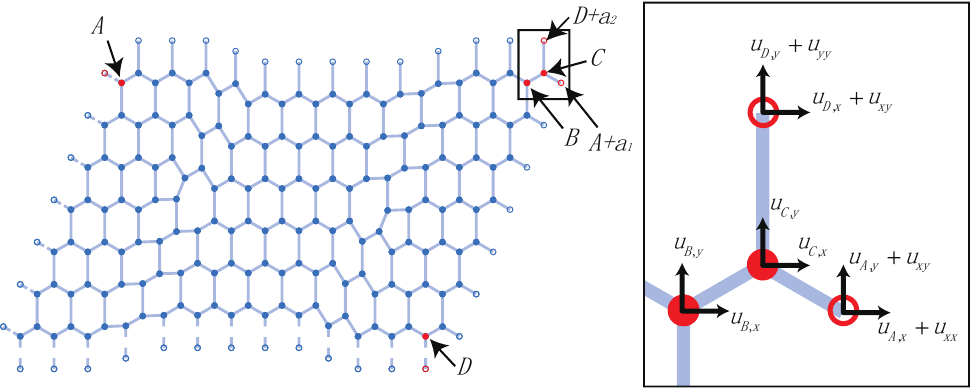}
\caption{(a) Schematic illustration of the augmented unit cell of the elliptically geared honeycomb lattice with a pair of opposite dislocations. (b) The displacements of sites $A$, $B$, $C$, and $D$, where the deformation of the unit cell (i.e., change of the primitive vectors) are marked by $u_{xx}$, $u_{xy}$, and $u_{yy}$.
}\label{S11}
\end{figure}

Finally, we show that Guest modes always arise in isostatic systems even if translational symmetry is broken by randomness or dislocations. 
As shown in Fig. \ref{S11}(a), an elliptically geared generalized honeycomb lattice is constructed to host a pair of opposite dislocations. As the coordination number is $z=3$ for every node, the lattice is still at the isostatic point. We connect the left, right boundaries, and the top, bottom boundaries to numerically impose PBCs. As such, we treat the frame in Fig. \ref{S11}(a) as an enlarged unit cell, whose primitive vectors are allowed to change nonlinearly and enable the geometric shearing of the whole structure. As marked by Fig. \ref{S11}, the change of the primitive vectors (i.e., the geometric change of the whole unit cell) is realized by introducing additional displacements $u_{xx}$, $u_{xy}$, and $u_{yy}$ on the nodes A and D as they travel along one primitive vector of this enlarged unit cell. Taking account of these additional degrees of freedom, the originally $N_{\rm DOF}\times N_{\rm DOF}$ compatibility matrix is now augmented to a $N_{\rm DOF}\times (N_{\rm DOF}+3)$ matrix, in which three extra null vectors correspond to two rigid-body translations and a mechanism shown in Fig. 4(b) of the main text. Notably, this mechanism emerges the nonlinear geometric deformation from Fig. 4(a) to Fig. 4(c) (i.e., experimentally verified from Fig. 4(d) to Fig. 4(f)) without disentangling the mechanical frame, enabling the topological phase transitions. It should be emphasized that although a pair of topological dislocations are included in our numerics for the periodic boundary conditions, the emergence of Guest mode is independent of the the boundary condition. Rather, it is determined by the isostaticity of the underlying mechanical frame. Thus, the Guest mode still arises and operates the lattice geometry if there is only one dislocation, as shown in Fig.4 of the main text.

\endwidetext

\end{document}